\documentclass[aps,
prd,
preprint,
nofootinbib,
bibnotes
]{revtex4-1}
\usepackage{amsmath,bbm,graphicx,hyperref,mathrsfs,slashed}
\usepackage[UTF8, scheme=plain]{ctex}
\begin{document}

\title{Beyond Symmetries : Anomalies in Transverse Ward--Takahashi Identities}
\author{Yi-Da Li}
\email{yd-li16@mails.tsinghua.edu.cn}
\affiliation{Department of Physics, Tsinghua University, Beijing 100084, P. R. China}

\author{Qing Wang}
\email[Corresponding author :~]{wangq@mail.tsinghua.edu.cn}
\affiliation{Department of Physics, Tsinghua University, Beijing 100084, P. R. China \\
Center for High Energy Physics, Tsinghua University, Beijing 100084, P. R. China}

\begin{abstract}
Anomalies in transverse Ward--Takahashi identities are studied, allowing discussion of the feasibility of anomalies arising in general non-symmetry Ward--Takahashi identities. We adopt the popular Fujikawa's method and rigorous dimensional renormalization to verify the existence of transverse anomalies to one-loop order and any loop order, respectively. The arbitrariness of coefficients of transverse anomalies is revealed, and a way out is also proposed after relating transverse anomalies to Schwinger terms and comparing symmetry and non-symmetry anomalies. Papers that claim the non-existence of transverse anomalies are reviewed to find anomalies hidden in their approaches. The role played by transverse anomalies is discussed.
\end{abstract}

\maketitle

\section{Introduction}\label{sec:intro}

There are always surprises in common quantization procedures let alone the quantization of relativistic fields, which is highly entangled with infinite degrees of freedom, with various divergences and anomalies revealing the power of quantum laws. We discuss such anomalies in the present paper. In various examples, anomalies break many symmetries and manifest as various anomalous Ward--Takahashi identities\cite{ward1950,takahashi1957} (WTIs), such as chiral\cite{adler-bardeen1969,jackiw1969,adler1969} and trace\cite{fujikawa1980,fujikawa2004} anomalies. To our knowledge, however, no paper has discussed anomalies in WTIs that do not stand for any symmetry. Our research on anomalies in the transverse Ward--Takahashi identity\cite{kondo1997,he2000,he2009,qin2013} (tWTI) which is not a symmetry WTI, has opened the door to non-symmetry anomalies. The content of non-symmetry WTI is much richer than that of symmetry WTIs, and the anomaly may be largely extended and more exposed to us such that we may see the nature of the anomaly more deeply. However, this requires far more examples of non-symmetry anomalies, apart from anomalies in tWTIs as discussed in this paper. Further support for non-symmetry anomalies must be left to further discoveries, with the present paper focusing on anomalies in tWTIs (referred to as transverse anomalies) only.

In dealing with transverse anomalies, we find that many methods applied to symmetry anomalies are entirely suitable for locating and analyzing a non-symmetry anomaly. As an example, if we use dimensional renormalization, anomalies in symmetry and non-symmetry WTIs may be treated on an equal footing because extra dimensional operators appear in equations of motion and any WTI that involves the use of equations of motion may acquire an anomaly because these extra dimensional operators can often be expanded using various four-dimensional operators that potentially include anomaly terms\cite{bonneau2,renormalization-collins}. Indeed, we find no difference in analyzing transverse anomalies when adopting dimensional renormalization in Sec.\ref{sec:derivation-DR} than when adopting a procedure to handle chiral and trace anomalies (as described in detail in \cite{bonneau2}).

Although dimensional renormalization already allows us to go to any order in perturbation theory, it's interesting and inspiring to look at some semi-classical one-loop methods in locating anomalies in WTIs. Fujikawa's elegant approach\cite{fujikawa1979,fujikawa2004} tells us that anomalies appear as long as we get non-trivial Jacobian factors by varying fields in the path integral when obtaining WTIs. It's therefore convenient to check whether a WTI has anomalies if we know how to get the WTI by varying fields, and this is the case for the tWTI. Equivalently\footnote{The equivalence of Fujikawa's method and the following canonical approach is proved in \cite{tsutsui1989}.}, we may locate anomalies in the canonical framework to avoid dealing with the explicit but somehow abstract path integral measure because it has no classical correspondence, and we do not even have to know how to get this WTI by varying fields. In the canonical framework without any consideration of renormalization, it is easy to conclude that an anomaly is a matter of definition of operators and that anomalies simply hide in equations of motion. We consider, for example, the case of the chiral anomaly\cite{satish1987,tsutsui1989}. Starting from the following Dyson--Schwinger equation, where derivatives should be outside the time ordered product (as is the convention used throughout the present paper), we have 
 \begin{equation}
\left<Ti\slashed{D}^x_{k\ell}\psi_\ell(x)\bar{\psi}_n(y)\right>=m\left<T\psi_k(x)\bar{\psi}_n(y)\right>+i\delta_{kn}\delta^4(x-y).
\label{eq:example}\end{equation}
In assembling the chiral anomaly, we contract $\left(\gamma_5\right)_{nk}$ on both sides of (\ref{eq:example}) and take the $y\rightarrow x$ limit, thus obtaining the familiar expression $\mathrm{tr}[\gamma_5]\delta^4(x-x)$, a signal for chiral anomaly\footnote{The minus sign is due to Fermi-statistics, and we drop the time-ordered symbol in the equal time limit.} :
\begin{equation}
\left<\bar{\psi}(x)\gamma_5i\slashed{\overrightarrow{D}}\psi(x)\right>=m\left<\bar{\psi}(x)\gamma_5\psi(x)\right>-i\mathrm{tr}[\gamma_5]\delta^4(x-x).
\label{eq:chiral1}\end{equation}
Similarly, we have (the Dyson--Schwinger equation for $\bar{\psi}$) 
\begin{equation}
\left<\bar{\psi}(x)i\slashed{\overleftarrow{D}}\gamma_5\psi(x)\right>=-m\left<\bar{\psi}(x)\gamma_5\psi(x)\right>+i\mathrm{tr}[\gamma_5]\delta^4(x-x),
\label{eq:chiral2}\end{equation}
The anomalous partial conservation equation for axial current $j_5^\mu\equiv\bar{\psi}\gamma^\mu\gamma_5\psi$ is then obtained ($\overrightarrow{D}_\mu=\overrightarrow{\partial}_\mu-igA_\mu,\overleftarrow{D}_\mu=\overleftarrow{\partial}_\mu+igA_\mu$) :
\begin{equation}\begin{aligned}
&\partial_\mu\left<j^\mu_5(x)\right>\\
=&-i\left<\bar{\psi}(x)i\slashed{\overleftarrow{D}}\gamma_5\psi(x)\right>+i\left<\bar{\psi}(x)\gamma_5i\slashed{\overrightarrow{D}}\psi(x)\right>\\
=&2im\left<\bar{\psi}(x)\gamma_5\psi(x)\right>+2\mathrm{tr}[\gamma_5]\delta^4(x-x).
\end{aligned}\label{eq:chiral anomaly}\end{equation}
This is exactly what Fujikawa\cite{fujikawa1979} obtained\footnote{Note that Fujikawa worked in Wick-rotated Euclidean space, thus an extra factor $i$ should be multiplied to our anomaly terms to restore his results. This is also true for the trace anomaly which we will talk about at once.} by calculating the transformation Jacobian of the path integral measure (before regularization). Additionally, it is easy to generalize to other anomalies such as the trace anomaly\footnote{\cite{trace1993} worked out trace anomaly of scalar field in curved spacetime, but in the present paper we talk about trace anomaly in quantum electrodynamics.} from conformal symmetry.  Starting from the following energy-momentum tensor (see e.g. (1.2) in \cite{tr1977}) :
\begin{equation}\begin{aligned}
\theta_{\mu\nu}=&\frac{1}{4}g_{\mu\nu}F_{\rho\sigma}F^{\rho\sigma}-F_{\mu\rho}F_{\nu}^{\ \rho}\\
&+\frac{i}{4}\left(\bar{\psi}\gamma_\mu \overrightarrow{D}_\nu\psi+\bar{\psi}\gamma_\nu \overrightarrow{D}_\mu\psi-\bar{\psi}\gamma_\mu \overleftarrow{D}_\nu\psi-\bar{\psi}\gamma_\nu \overleftarrow{D}_\mu\psi\right),
\end{aligned}\end{equation}
and contract (\ref{eq:example}) with $\delta_{kn}$, i.e.,
\begin{equation}
\left<\bar{\psi}(x)i\slashed{\overrightarrow{D}}\psi(x)\right>=m\left<\bar{\psi}(x)\psi(x)\right>-i\mathrm{tr}\left[\mathbbm{1}\right]\delta^4(x-x).
\end{equation}
Together with that of $\bar{\psi}$ :
\begin{equation}
\left<\bar{\psi}(x)i\slashed{\overleftarrow{D}}\psi(x)\right>=-m\left<\bar{\psi}(x)\psi(x)\right>+i\mathrm{tr}\left[\mathbbm{1}\right]\delta^4(x-x),
\end{equation}
we have 
\begin{equation}
g^{\mu\nu}\left<\theta_{\mu\nu}(x)\right>=m\left<\bar{\psi}(x)\psi(x)\right>-i\mathrm{tr}\left[\mathbbm{1}\right]\delta^4(x-x).
\end{equation}
Again, this reproduces what Fujikawa obtained by his method\cite{fujikawa1981} (before regularization).

Clearly, the important step in getting the chiral and trace anomaly explicitly is to define $\bar{\psi}(x)\gamma_5i\slashed{\overrightarrow{D}}\psi(x)$ and $\bar{\psi}(x)i\slashed{\overrightarrow{D}}\psi(x)$ to be $\lim_{y\rightarrow x}T\bar{\psi}(y)\gamma_5i\slashed{\overrightarrow{D}}_x\psi(x)$ and $\lim_{y\rightarrow x}T\bar{\psi}(y)i\slashed{\overrightarrow{D}}_x\psi(x)$. If we use the naive definition that $\bar{\psi}(x)\gamma_5i\slashed{\overrightarrow{D}}\psi(x)\equiv\lim_{y\rightarrow x}\bar{\psi}(y)\gamma_5i\slashed{\overrightarrow{D}}_x\psi(x)$ and $\bar{\psi}(x)i\slashed{\overrightarrow{D}}\psi(x)\equiv\lim_{y\rightarrow x}\bar{\psi}(y)i\slashed{\overrightarrow{D}}_x\psi(x)$, there can not be any anomalous terms in the chiral and trace WTI. According to the above argument, as long as equations of motion are used in derivations of a WTI and the time-ordered product definition of operators is taken, an anomaly in the form of singular contact terms like $\mathrm{tr}[\gamma_5]\delta^4(x-x)$ and $\mathrm{tr}[\mathbbm{1}]\delta^4(x-x)$ may appear\footnote{Of course, this is only established on one-loop order and some specific regularization schemes such as $\zeta$ function regularization discussed in App.\ref{app:zeta}.}. This helps us greatly to anticipate possible anomalies in new WTIs --- not necessarily one that stands for some symmetry --- before resorting to rigorous all-order methods, such as dimensional renormalization.

The remainder of the paper is organized as follows. We first briefly review the tWTI in Sec.\ref{sec:review} and then derive the tWTI in Fujikawa's paradigm (one-loop order) in Sec.\ref{sec:derivation-fujikawa} to obtain intuitive ideas on transverse anomalies.  We next present a rigorous any-loop order analysis of the tWTI in dimensional renormalization in Sec.\ref{sec:derivation-DR}. In Sec.\ref{sec:ST}, we discuss the connection between transverse anomalies and Schwinger terms on the basis of Sec.\ref{sec:derivation-DR}. It has been shown many times that the naive tWTI (i.e., without transverse anomalies) is correct on one-loop order, and we make comments in Sec.\ref{sec:comments}  and App.\ref{app:pv} and App.\ref{app:ps} relating to picking up hidden transverse anomalies in those approaches. Symmetry and non-symmetry anomalies are then compared in Sec.\ref{sec:comparison}. We conclude the paper in Sec.\ref{sec:conclusion}. It is worth emphasizing that throughout the paper except in Sec.\ref{sec:derivation-fujikawa} and where one-loop is indicated explicitly, we work with the accuracy to any loop order in perturbation theory.

In this paper, the space-time metric is $g^{\mu\nu}=\mathrm{diag}(+,-,-,-)$. $\gamma_5\equiv i\gamma^0\gamma^1\gamma^2\gamma^3$ and $\epsilon^{0123}=+1$. We define  $\sigma^{\mu\nu}\equiv\frac{i}{2}[\gamma^\mu,\gamma^\nu]$.

\section{Review of $\mathbf{tWTIs}$}\label{sec:review}

Vector and axial vector tWTIs have been proposed\cite{kondo1997,he2000,he2009,qin2013} in Abelian and non-Abelian cases. In this paper, we focus on the Abelian case only while the non-Abelian generalization is presented in App.\ref{app:zeta}(see (\ref{eq:non-abelian})) but is to be investigated in detail elsewhere.

The Abelian vector tWTI\footnote{In most cases, this is the meaning of the tWTI.} is often presented as\cite{he2009,qin2013}(without anomaly; the generalization that includes more $\psi\left(y_i\right)$ and $\bar{\psi}\left(z_j\right)$ or $A^{\mu_k}\left(u_k\right)$ is obvious)
\begin{equation}\begin{aligned}
&\partial^\mu_x\left<Tj^\nu(x)\psi(y)\bar{\psi}(z)\right>-\partial^\nu_x\left<Tj^\mu(x)\psi(y)\bar{\psi}(z)\right>\\
=&i\sigma^{\mu\nu}\left<T\psi(y)\bar{\psi}(z)\right>\delta^4(x-y)+i\left<T\psi(y)\bar{\psi}(z)\right>\sigma^{\mu\nu}\delta^4(x-z)\\
&+i\epsilon^{\mu\nu\rho\sigma}\left(\partial^x_\rho-\partial^{x'}_\rho\right)\left<T\bar{\psi}(x')\gamma_\sigma\gamma_5e^{ig\int_x^{x'}dy\cdot A}\psi(x)\psi(y)\bar{\psi}(z)\right>_{x'\rightarrow x}\\
&+2m\left<T \bar{\psi}(x)\sigma^{\mu\nu}\psi(x)\psi(y)\bar{\psi}(z)\right>,
\end{aligned}\label{eq:tWTI without anomaly}\end{equation}
while the axial vector tWTI is 
\begin{equation}\begin{aligned}
&\partial^\mu_x\left<Tj^\nu_5(x)\psi(y)\bar{\psi}(z)\right>-\partial^\nu_x\left<Tj^\mu_5(x)\psi(y)\bar{\psi}(z)\right>\\
=&i\sigma^{\mu\nu}\gamma_5\left<T\psi(y)\bar{\psi}(z)\right>\delta^4(x-y)-i\left<T\psi(y)\bar{\psi}(z)\right>\sigma^{\mu\nu}\gamma_5\delta^4(x-z)\\
&+i\epsilon^{\mu\nu\rho\sigma}\left(\partial^x_\rho-\partial^{x'}_\rho\right)\left<T\bar{\psi}(x')\gamma_\sigma e^{ig\int_x^{x'}dy\cdot A}\psi(x)\psi(y)\bar{\psi}(z)\right>_{x'\rightarrow x}.
\end{aligned}\label{eq:axial tWTI}\end{equation}

Before starting, it is necessary to reduce both (\ref{eq:tWTI without anomaly}) and (\ref{eq:axial tWTI}) to simpler forms. The apparently non-local expression of $\lim\limits_{x'\rightarrow x}i\epsilon^{\mu\nu\rho\sigma}\left(\partial^x_\rho-\partial^{x'}_\rho\right)\bar{\psi}(x')\gamma_\sigma\gamma_5e^{ig\int_x^{x'}dy\cdot A}\psi(x)$ is suitable for Fourier transformations\cite{he2009} but a little confusing because the factor $e^{ig\int_x^{x'}dy\cdot A}$ is not used in this paper. The way out is to simply work out this limit first\footnote{In fact, the original expression is just the \emph{result} of the limit, so any question about the interchange of limits is not of concern here, as easily seen in Sec.\ref{sec:derivation-fujikawa} and Sec.\ref{sec:derivation-DR}.} :
\begin{equation}
\lim\limits_{x'\rightarrow x}i\epsilon^{\mu\nu\rho\sigma}\left(\partial^x_\rho-\partial^{x'}_\rho\right)\bar{\psi}(x')\gamma_\sigma\gamma_5e^{ig\int_x^{x'}dy\cdot A}\psi(x)=2\epsilon^{\mu\nu\rho\sigma}\bar{\psi}(x)\gamma_\sigma\gamma_5iD_\rho\psi(x),
\end{equation}
where $D_\rho=\frac{1}{2}\left(\overrightarrow{\partial}_\rho-\overleftarrow{\partial}_\rho\right)-igA_\rho$. We hereafter use $2\epsilon^{\mu\nu\rho\sigma}\bar{\psi}(x)\gamma_\sigma\gamma_5iD_\rho\psi(x)$ rather than the non-local limit (the same as that in (\ref{eq:axial tWTI}) with $\gamma_\sigma\gamma_5\rightarrow\gamma_\sigma$).

We return to the tWTI. (\ref{eq:tWTI without anomaly}) and (\ref{eq:axial tWTI}) are not conservation equations for any currents because transformations in (\ref{eq:var-tWTI}) leading to tWTIs with $\alpha(x)={Const.}$ do not leave the Lagrangian or action invariant, even with $m\rightarrow0$. Therefore, the tWTI is a proper example that illustrates the richness of anomalies beyond the scope of quantum obstacles to classical symmetries.

\section{Heuristic Derivation of Transverse Anomalies using Fujikawa's Method}\label{sec:derivation-fujikawa}

In this section, we make use of Fujikawa's method for the path integral measure to obtain some intuitive pictures of transverse anomalies. It is known\cite{fujikawa2004} that Fujikawa's original method is correct only in the sense of the background field approximation, and we thus treat $A_\mu(x)$ as a background electromagnetic potential and the following Lagrangian should be sufficient in this section (i.e., there is no need for renormalization at the moment). The Lagrangian is
\begin{equation}
\mathscr{L}_{\mathrm{BG}}=\bar{\psi}\left(\frac{i}{2}\overleftrightarrow{\slashed{\partial}}-ig_0\slashed{A}-m_0\right)\psi.
\label{eq:lag-BG}\end{equation}
The partition function $Z\left[\eta,\bar{\eta},A\right]$ is simply :
\begin{equation}
Z\left[\eta,\bar{\eta},A\right]=\int\left[d\psi d\bar{\psi}\right]e^{i\int d^4x\left(\mathscr{L}_{\mathrm{BG}}(x)+\bar{\psi}(x)\eta(x)+\bar{\eta}(x)\psi(x)\right)}.
\label{eq:partition-func}\end{equation}
In the absence of the dynamics of $A_\mu$, (\ref{eq:partition-func}) is simply a one-loop approximation of quantum electrodynamics (QED).

We apply the field variation 
\begin{equation}\begin{aligned}
\delta\psi(x)=&\frac{1}{4}\alpha(x)\epsilon_{\mu\nu}\sigma^{\mu\nu}\psi(x),\\
\delta\bar{\psi}(x)=&\frac{1}{4}\alpha(x)\epsilon_{\mu\nu}\bar{\psi}(x)\sigma^{\mu\nu}.
\end{aligned}\label{eq:var-tWTI}\end{equation}
and include its non-trivial Jacobian\footnote{See (\ref{eq:jacobian}) for a detailed derivation.} and thus obtain the desired tWTI :
\begin{equation}\begin{aligned}
&\partial^\mu\left<j^\nu(x)\right>_A-\partial^\nu\left<j^\mu(x)\right>_A\\
=&2\epsilon^{\mu\nu\rho\sigma}\left<\bar{\psi}(x)\gamma_\sigma\gamma_5iD_\rho\psi(x)\right>_A+2m\left<\bar{\psi}(x)\sigma^{\mu\nu}\psi(x)\right>_A\\
&-\left<\bar{\psi}(x)\right>_A\sigma^{\mu\nu}\eta(x)-\bar{\eta}(x)\sigma^{\mu\nu}\left<\psi(x)\right>_A-2i\mathrm{tr}\left[\sigma^{\mu\nu}\right]\delta^4(x-x).
\end{aligned}\label{tWTI-BG1}\end{equation}
We thus focus on the non-trivial Jacobian of the path integral measure, i.e., $-2i\mathrm{tr}\left[\sigma^{\mu\nu}\right]\delta^4(x-x)$ in (\ref{tWTI-BG1}). In contrast, the field variation leading to the axial tWTI,
\begin{equation}\begin{aligned}
\delta\psi(x)=&+\frac{1}{4}\alpha(x)\epsilon_{\mu\nu}\sigma^{\mu\nu}\gamma_5\psi(x),\\
\delta\bar{\psi}(x)=&-\frac{1}{4}\alpha(x)\epsilon_{\mu\nu}\bar{\psi}(x)\sigma^{\mu\nu}\gamma_5,
\end{aligned}\label{eq:var-atWTI}\end{equation}
acquires a vanishing Jacobian factor owing to the different signs of $\delta\psi$ and $\delta\bar{\psi}$ and thus contributes no anomalies. From now on, we will consider the vector tWTI only.

Proceeding with Fujikawa's original regularization method (i.e., with the regulator $e^{-\slashed{D}^2/\Lambda^2}$), we get the divergent result (see App.\ref{app:zeta} for details)
\begin{equation}
-2i\mathrm{tr}\left[\sigma^{\mu\nu}\right]\delta^4(x-x)\overset{\Lambda}{=}\frac{g_0\Lambda^2}{2\pi^2}F^{\mu\nu}(x)-\frac{g_0}{12\pi^2}\partial_\rho\partial^\rho F^{\mu\nu}(x).
\label{eq:fujikawa}\end{equation}
However, it is known\cite{peskin} that quadratic divergence in QED corresponds to photon mass and thus must be discarded. An elegant way to do this is to employ the $\zeta$ function regularization (which effectively turns $\Lambda^2$ to $-m^2_0$) and thus obtain
\begin{equation}
-2i\mathrm{tr}\left[\sigma^{\mu\nu}\right]\delta^4(x-x)\overset{\zeta}{=}-\frac{g_0m_0^2}{2\pi^2}F^{\mu\nu}(x)-\frac{g_0}{12\pi^2}\partial_\rho\partial^\rho F^{\mu\nu}(x).
\end{equation}
The final result of the tWTI is thus 
\begin{equation}\begin{aligned}
&\partial^\mu\left<j^\nu(x)\right>_A-\partial^\nu\left<j^\mu(x)\right>_A\\
=&2\epsilon^{\mu\nu\rho\sigma}\left<\bar{\psi}(x)\gamma_\sigma\gamma_5iD_\rho\psi(x)\right>_A+2m\left<\bar{\psi}(x)\sigma^{\mu\nu}\psi(x)\right>_A\\
&-\left<\bar{\psi}(x)\right>_A\sigma^{\mu\nu}\eta(x)-\bar{\eta}(x)\sigma^{\mu\nu}\left<\psi(x)\right>_A\\
&-\frac{g_0m_0^2}{2\pi^2}F^{\mu\nu}(x)-\frac{g_0}{12\pi^2}\partial_\rho\partial^\rho F^{\mu\nu}(x).
\end{aligned}\label{tWTI-BG2}\end{equation}

\begin{figure}[ht]
\centering
\includegraphics[width=6cm]{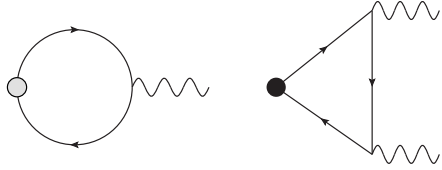}
\caption{(Divergent) one-loop diagrams for operators in the tWTI and chiral WTI (gray vertex for those in the tWTI and black vertex for those in the chiral WTI)}
\label{fig:BG-1}
\end{figure}

At this point, however, it is emphasized that$\left<j^\mu(x)\right>_A$, $\epsilon^{\mu\nu\rho\sigma}\left<\bar{\psi}(x)\gamma_\sigma\gamma_5iD_\rho\psi(x)\right>_A$ and $\left<\bar{\psi}(x)\sigma^{\mu\nu}\psi(x)\right>_A$ in (\ref{tWTI-BG2}) are not well defined (even in the background field approximation (i.e., 1-loop order)) owing to the divergence of loop diagrams with only two vertices even after imposing gauge invariance in the external photon leg. This is unlike the case of the chiral WTI, where degrees of divergence of triangle diagrams are largely decreased by both an additional internal fermion propagator and gauge invariance in two external photon legs. See Fig.\ref{fig:BG-1} .

It is the divergence in these loop diagrams that makes the anomaly terms ambiguous because\footnote{The authors are in debt to the referee for pointing this out.} they maybe only counterterms of $\epsilon^{\mu\nu\rho\sigma}\left<\bar{\psi}(x)\gamma_\sigma\gamma_5iD_\rho\psi(x)\right>_A$ and $\left<\bar{\psi}(x)\sigma^{\mu\nu}\psi(x)\right>_A$. However, we will see that the anomaly terms survive even after renormalization. A rigorous analysis using dimensional renormalization is presented in the next section. Of course, the gauge fields $A_\mu$ present are not treated as an external source.

\section{Full Analysis in Dimensional Renormalization}\label{sec:derivation-DR}

For simplicity and clarity, the modified minimal subtraction ($\overline{\mathrm{MS}}$) is used in this section. We first specify the effective Lagrangian\cite{bonneau2} (i.e., without infinite counterterms) :
\begin{equation}\begin{aligned}
\mathscr{L}_{\mathrm{EFF}}=&-\frac{1}{4}F_{\mu\nu}F^{\mu\nu}+\bar{\psi}\left(i\slashed{D}-m\right)\psi\\
&+\frac{\lambda^2}{2}A_\mu A^\mu-\frac{1}{2\xi}\left(\partial_\mu A^\mu\right)^2.
\end{aligned}\end{equation}
Here the gauge fixing term is as usual, and a photon mass term is added to regularize infrared divergences\cite{bonneau2}. Now that $A_\mu$ is dynamical, our analysis can be extended to all orders by virtue of dimensional renormalization\cite{dim,breit-maison,bonneau1,renormalization-collins}.

As figured out in a series of papers\cite{breit-maison,bonneau1,bonneau2}, anomalies in dimensional renormalization arise from extra dimensional objects like\footnote{We here use the same convention as used by Collins\cite{renormalization-collins}, where objects with a bar, such as $\bar{\gamma}^\mu$, are genuinely four-dimensional things (but not to be confused with the bar in $\bar{\psi}$, which indicates pseudo-Hermitian conjugation), objects with a hat, such as $\hat{g}^{\mu\nu}$, exist in extra dimensions, and objects without special labels, such as $p^\mu=\bar{p}^\mu+\hat{p}^\mu$, are complete $d$-dimensional entities.} $\hat{g}^\mu_\mu=d-4$ and $\left\{\gamma^\mu,\gamma_5\right\}=2\hat{\gamma}^\mu\gamma_5$ which correspond respectively to trace and chiral anomalies, as do transverse anomalies. Using the normal product formalism\cite{renormalization-collins,bonneau1,bonneau2} in dimensional renormalization, we can easily derive a prototype of the tWTI :
\begin{equation}\begin{aligned}
&\partial^\mu N\left[j^\nu\right]-\partial^\nu N\left[j^\mu\right]=-\frac{i}{2}\partial_\rho N\left[\bar{\psi}\left[\gamma^\rho,\bar{\sigma}^{\mu\nu}\right]\psi\right]\\
=&2N\left[\epsilon^{\mu\nu\rho\sigma}\bar{\psi}\gamma_\sigma\gamma_5 iD_\rho\psi\right]+2mN\left[\bar{\psi}\bar{\sigma}^{\mu\nu}\psi\right]-2N\left[\bar{\psi}\bar{\sigma}^{\mu\nu}\hat{\gamma}^\rho iD_\rho\psi\right]\\
&+N\left[\bar{\psi}\bar{\sigma}^{\mu\nu}\left(i\overrightarrow{\slashed{D}}-m\right)\psi\right]+N\left[\bar{\psi}\left(-i\overleftarrow{\slashed{D}}-m\right)\bar{\sigma}^{\mu\nu}\psi\right].
\end{aligned}\label{tWTI-DR1}\end{equation}

Formally vanished $N\left[\bar{\psi}\bar{\sigma}^{\mu\nu}\left(i\overrightarrow{\slashed{D}}-m\right)\psi\right]$ and $N\left[\bar{\psi}\left(-i\overleftarrow{\slashed{D}}-m\right)\bar{\sigma}^{\mu\nu}\psi\right]$ obtained using equations of motion are sources of contact terms when inserted into Green's functions\cite{renormalization-collins}, and we thus only need consider the possible anomaly term $N\left[\bar{\psi}\bar{\sigma}^{\mu\nu}\hat{\gamma}^\rho iD_\rho\psi\right]$. Although $N\left[\bar{\psi}\bar{\sigma}^{\mu\nu}\hat{\gamma}^\rho iD_\rho\psi\right]$ has a vanishing tree diagram, its loop diagrams are not zero because $\overleftrightarrow{\partial}_\rho$ \emph{inside} the operator drops loop momenta off\footnote{Extra dimensional loop momenta must not be taken to be zero before carrying out loop integrals, in contrast with the case for external momenta.} and thus has non-zero contributions. Operators with evanescent vertices\cite{renormalization-collins} like this are simply origins of anomalies in dimensional renormalization. Thanks to the existence of Zimmermann-like identities in dimensional renormalization\cite{bonneau1,bonneau2}, transverse anomalies that are also of the form $N\left[\hat{g}_{\mu\nu}\mathcal{O}^{\mu\nu}\right]$ (just as trace and chiral anomalies which were given by Bonneau\cite{bonneau2}) because
\begin{equation}
N\left[\bar{\psi}\bar{\sigma}^{\mu\nu}\hat{\gamma}^\rho iD_\rho\psi\right]\equiv N\left[\hat{g}_{\rho\sigma}\left(\bar{\psi}\bar{\sigma}^{\mu\nu}\gamma^\rho iD^\sigma\psi\right)\right]
\end{equation}
can be reduced to usual operators in physical dimensions :
\begin{equation}\begin{aligned}
N\left[\bar{\psi}\bar{\sigma}^{\mu\nu}\hat{\gamma}^\rho iD_\rho\psi\right]=&aN\left[\bar{\psi}\bar{\sigma}^{\mu\nu}\hat{\gamma}^\rho iD_\rho\psi\right]+b'\left(\partial^\mu N\left[j^\nu\right]-\partial^\nu N\left[j^\mu\right]\right)\\
&+c'N\left[\epsilon^{\mu\nu\rho\sigma}\bar{\psi}\gamma_\sigma\gamma_5 iD_\rho\psi\right]+f'mN\left[\bar{\psi}\bar{\sigma}^{\mu\nu}\psi\right]\\
&+r'\partial^\rho\partial_\rho F^{\mu\nu}+s'm^2F^{\mu\nu}.
\end{aligned}\label{eq:evan-1}\end{equation}
For simplicity, we omit the normal product symbol for a single renormalized field $F^{\mu\nu}$. Similar to the case of trace and chiral anomalies\cite{bonneau2}, all the above coefficients can be obtained from residues of the simple pole (r.s.p.) at $4-d=0$ of the overall divergence of specific Green's functions of $N\left[\check{g}_{\rho\sigma}\left(\bar{\psi}\bar{\sigma}^{\mu\nu}\gamma^\rho iD^\sigma\psi\right)\right]$, where $\check{g}_{\rho\sigma}$ defined by Bonneau\cite{bonneau2} is roughly $\hat{g}_{\rho\sigma}/(d-4)$ but $1/(d-4)$ therein is not included in the Laurent expansion when determining counterterms\cite{bonneau2}. The results are as follows\footnote{$\tilde{\psi}(p)\equiv\int d^4x\ e^{ip\cdot x}\psi(x),\tilde{\bar{\psi}}(q)\equiv\int d^4x\ e^{-iq\cdot x}\bar{\psi}(x),\tilde{A}^\rho(k)\equiv\int d^4x\ e^{ik\cdot x}A^\rho(x).$} (here, a line over a Green's function indicates the overall divergence (i.e., the counterterm obtained by contracting the whole 1PI diagram to a single vertex) has not been subtracted). We have
\begin{equation}\begin{aligned}
a=&\frac{1}{48(4-d)}\mathrm{tr}\left\{\mathrm{r.s.p.}\frac{\partial}{\partial p_\rho}\overline{\left<TN\left[\check{g}_{\alpha\beta}\left(\bar{\psi}\bar{\sigma}^{\mu\nu}\gamma^\alpha iD^\beta\psi\right)\right]\tilde{\psi}\left(\frac{1}{2}p\right)\tilde{\bar{\psi}}\left(\frac{1}{2}p\right)\right>^{\mathrm{prop}}}|_{\check{g},p=0}\times\bar{\sigma}^{\mu\nu}\hat{\gamma}_\rho\right\},\\
b'=&\frac{1}{-96i}\mathrm{tr}\left\{\mathrm{r.s.p.}\frac{\partial}{\partial p_\rho}\overline{\left<TN\left[\check{g}_{\alpha\beta}\left(\bar{\psi}\bar{\sigma}^{\mu\nu}\gamma^\alpha iD^\beta\psi\right)\right]\tilde{\psi}\left(\frac{1}{2}p\right)\tilde{\bar{\psi}}\left(-\frac{1}{2}p\right)\right>^{\mathrm{prop}}}|_{\check{g},p=0}\times\left(\bar{g}_{\mu\rho}\bar{\gamma}_\nu-\bar{g}_{\nu\rho}\bar{\gamma}_\mu\right)\right\},\\
c'=&\frac{1}{-96}\mathrm{tr}\left\{\mathrm{r.s.p.}\frac{\partial}{\partial p_\rho}\overline{\left<TN\left[\check{g}_{\alpha\beta}\left(\bar{\psi}\bar{\sigma}^{\mu\nu}\gamma^\alpha iD^\beta\psi\right)\right]\tilde{\psi}\left(\frac{1}{2}p\right)\tilde{\bar{\psi}}\left(\frac{1}{2}p\right)\right>^{\mathrm{prop}}}|_{\check{g},p=0}\times\epsilon_{\mu\nu\rho\sigma}\gamma^\sigma\gamma_5\right\},\\
f'=&\frac{1}{-48m}\mathrm{tr}\left\{\mathrm{r.s.p.}\overline{\left<TN\left[\check{g}_{\alpha\beta}\left(\bar{\psi}\bar{\sigma}^{\mu\nu}\gamma^\alpha iD^\beta\psi\right)\right]\tilde{\psi}(0)\tilde{\bar{\psi}}(0)\right>^{\mathrm{prop}}}|_{\check{g}=0}\times\bar{\sigma}_{\mu\nu}\right\},\\
r'=&\frac{1}{-288i}\frac{\partial}{\partial q_\delta}\frac{\partial}{\partial q^\delta}\frac{\partial}{\partial q_\sigma}\mathrm{r.s.p.}\overline{\left<TN\left[\check{g}_{\alpha\beta}\left(\bar{\psi}\bar{\sigma}^{\mu\nu}\gamma^\alpha iD^\beta\psi\right)\right]\tilde{A}^\rho(q)\right>^{\mathrm{prop}}}|_{\check{g},q=0}\times\left(\bar{g}_{\mu\rho}\bar{g}_{\nu\sigma}-\bar{g}_{\mu\sigma}\bar{g}_{\nu\rho}\right),\\
s'=&\frac{1}{-24im^2}\frac{\partial}{\partial q_\sigma}\mathrm{r.s.p.}\overline{\left<TN\left[\check{g}_{\alpha\beta}\left(\bar{\psi}\bar{\sigma}^{\mu\nu}\gamma^\alpha iD^\beta\psi\right)\right]\tilde{A}^\rho(q)\right>^{\mathrm{prop}}}_{\check{g},q=0}\times\left(\bar{g}_{\mu\sigma}\bar{g}_{\nu\rho}-\bar{g}_{\mu\rho}\bar{g}_{\nu\sigma}\right).
\end{aligned}\label{eq:coeffi-DR}\end{equation}
Finally, the tWTI is
\begin{equation}\begin{aligned}
&\left(1-2b\right)\left(\partial^\mu N\left[j^\nu\right]-\partial^\nu N\left[j^\mu\right]\right)\\
=&2\left(1-c\right)N\left[\epsilon^{\mu\nu\rho\sigma}\bar{\psi}\gamma_\sigma\gamma_5 iD_\rho\psi\right]+2m\left(1-f\right)N\left[\bar{\psi}\bar{\sigma}^{\mu\nu}\psi\right]\\
&-2r\partial^\rho\partial_\rho F^{\mu\nu}-2sm^2F^{\mu\nu}\\
&+N\left[\bar{\psi}\bar{\sigma}^{\mu\nu}\left(i\overrightarrow{\slashed{D}}-m\right)\psi\right]+N\left[\bar{\psi}\left(-i\overleftarrow{\slashed{D}}-m\right)\bar{\sigma}^{\mu\nu}\psi\right],
\end{aligned}\label{tWTI-DR2}\end{equation}
where $x=x'/(1-a)$ for $x=b,c,f,r,s$ and their one-loop values are given in App.\ref{app:dim}.
 
In QED, it is necessary to ensure that all coefficients are gauge invariant and, in particular, we should focus on coefficients of transverse anomalies, namely $-2r$ and $-2sm^2$. Fortunately, following the arguments made by Bonneau about gauge invariance of the chiral anomaly\cite{bonneau2}, mainly\footnote{Conclusions in \cite{bonneau2} are so general that nothing essential needs modifications.} (B.10), Lemma 3, and Fig.3 in \cite{bonneau2}, it is almost trivial to see that $b,c,fm,r,sm^2$ are all gauge invariant, i.e., independent of $\xi$. This gauge independence is also briefly discussed in App.\ref{app:dim}.

It is now clear that transverse anomalies exists even after renormalization. Of course, with the presence of the four-dimensional operator $N\left[\epsilon^{\mu\nu\rho\sigma}\bar{\psi}\gamma_\sigma\gamma_5 iD_\rho\psi\right]$ and three-dimensional operator $N\left[\bar{\psi}\bar{\sigma}^{\mu\nu}\psi\right]$, transverse anomalies (whose dimensions are respectively 4 and 2) may be absorbed into these operators as finite counterterms and thus be rendered dependent on the renormalization schemes \footnote{We thank the referee for making this point.}. We may represent the most general form of the tWTI as :
\begin{equation}\begin{aligned}
&\left(1-2b\right)\left(\partial^\mu N\left[j^\nu\right]-\partial^\nu N\left[j^\mu\right]\right)\\
=&2\left(1-c\right)\tilde{N}\left[\epsilon^{\mu\nu\rho\sigma}\bar{\psi}\gamma_\sigma\gamma_5 iD_\rho\psi\right]+2m\left(1-f\right)\tilde{N}\left[\bar{\psi}\bar{\sigma}^{\mu\nu}\psi\right]\\
&-2\tilde{r}\partial^\rho\partial_\rho F^{\mu\nu}-2\tilde{s}m^2F^{\mu\nu}\\
&+N\left[\bar{\psi}\bar{\sigma}^{\mu\nu}\left(i\overrightarrow{\slashed{D}}-m\right)\psi\right]+N\left[\bar{\psi}\left(-i\overleftarrow{\slashed{D}}-m\right)\bar{\sigma}^{\mu\nu}\psi\right].
\end{aligned}\label{tWTI-DR3}\end{equation}
Here, the operators with a tilde are simply linear combinations of the original operators and transverse anomalies :
\begin{equation}\begin{aligned}
2\left(1-c\right)\tilde{N}\left[\epsilon^{\mu\nu\rho\sigma}\bar{\psi}\gamma_\sigma\gamma_5 iD_\rho\psi\right]\equiv&2\left(1-c\right)N\left[\epsilon^{\mu\nu\rho\sigma}\bar{\psi}\gamma_\sigma\gamma_5 iD_\rho\psi\right]-2\left(r-\tilde{r}\right)\partial_\rho\partial^\rho F^{\mu\nu}-2\tilde{\alpha}\left(s-\tilde{s}\right)m^2F^{\mu\nu},\\
2m\left(1-f\right)\tilde{N}\left[\bar{\psi}\bar{\sigma}^{\mu\nu}\psi\right]\equiv&2m\left(1-f\right)N\left[\bar{\psi}\bar{\sigma}^{\mu\nu}\psi\right]-2\left(1-\tilde{\alpha}\right)\left(s-\tilde{s}\right)m^2F^{\mu\nu},
\end{aligned}\end{equation}
and $\tilde{r},\tilde{s},\tilde{\alpha}$ denote arbitrary real numbers of order $\mathcal{O}(g^2)$.

We will talk more generally about this arbitrariness in Sec.\ref{sec:comparison} by comparing with chiral and trace anomalies. However, we suggest a way of fixing the coefficient $\tilde{r}$ making use of Schwinger terms in the next section.

\section{Transverse anomalies and Schwinger terms}\label{sec:ST}

The Green's function version of (\ref{tWTI-DR3}), together with $\psi(y)\bar{\psi}(z)A^\rho(u)$, is :
\begin{equation}\begin{aligned}
&\left(1-2b\right)\left(\partial^\mu_x \left<TN\left[j^\nu\right](x)\psi(y)\bar{\psi}(z)A^\rho(u)\right>-\partial^\nu_x \left<TN\left[j^\mu\right](x)\psi(y)\bar{\psi}(z)A^\rho(u)\right>\right)\\
=&2\left(1-c\right)\left<T\tilde{N}\left[\epsilon^{\mu\nu\rho\sigma}\bar{\psi}\gamma_\sigma\gamma_5iD_\rho\psi\right](x)\psi(y)\bar{\psi}(z)A^\rho(u)\right>\\
&+2m\left(1-f\right)\left<T\tilde{N}\left[\bar{\psi}\bar{\sigma}^{\mu\nu}\psi\right](x)\psi(y)\bar{\psi}(z)A^\rho(u)\right>\\
&-2\tilde{r}\left<T\partial_x^\rho\partial_\rho^x F^{\mu\nu}(x)\psi(y)\bar{\psi}(z)A^\rho(u)\right>-2sm^2\left<TF^{\mu\nu}(x)\psi(y)\bar{\psi}(z)A^\rho(u)\right>\\
&+i\bar{\sigma}^{\mu\nu}\left<T\psi(y)\bar{\psi}(z)A^\rho(u)\right>\delta^4(x-y)+i\left<T\psi(y)\bar{\psi}(z)A^\rho(u)\right>\bar{\sigma}^{\mu\nu}\delta^4(x-z).
\end{aligned}\end{equation}

The crucial observation is to note the equation of motion for photon field $A^\rho(u)$ :
\begin{equation}\begin{aligned}
&\partial_\mu^x\left<TF^{\mu\nu}(x) \psi(y)\bar{\psi}(z)A^\rho(u)\right>+\frac{1}{\xi}\partial^\nu_x\left<T\partial_\mu^xA^\mu(x)\psi(y)\bar{\psi}(z)A^\rho(u)\right>\\
=&-g\left<Tj^\nu(x)\psi(y)\bar{\psi}(z)A^\rho(u)\right>+ig^{\nu\rho}\left<T\psi(y)\bar{\psi}(z)\right>\delta^4(x-u).
\end{aligned}\end{equation}
Taking the equation of motion together with the Bianchi identity\cite{peskin} :
\begin{equation}
\partial^\mu F^{\nu\rho}+\partial^\nu F^{\rho\mu}+\partial^\rho F^{\mu\nu}=0,
\end{equation}
it is easy to get
\begin{equation}\begin{aligned}
&\left(1-2b-2g\tilde{r}\right)\left(\partial^\mu_x \left<TN\left[j^\nu\right](x)\psi(y)\bar{\psi}(z)A^\rho(u)\right>-\partial^\nu_x \left<TN\left[j^\mu\right](x)\psi(y)\bar{\psi}(z)A^\rho(u)\right>\right)\\
=&2\left(1-c\right)\left<T\tilde{N}\left[\epsilon^{\mu\nu\rho\sigma}\bar{\psi}\gamma_\sigma\gamma_5iD_\rho\psi\right](x)\psi(y)\bar{\psi}(z)A^\rho(u)\right>\\
&+2m\left(1-f\right)\left<T\tilde{N}\left[\bar{\psi}\bar{\sigma}^{\mu\nu}\psi\right](x)\psi(y)\bar{\psi}(z)A^\rho(u)\right>\\
&-2i\tilde{r}\left<T\psi(y)\bar{\psi}(z)\right>\left(\partial^\mu_x g^{\nu\rho}-\partial^\nu_x g^{\mu\rho}\right)\delta^4(x-u)-2\tilde{s}m^2\left<TF^{\mu\nu}(x)\psi(y)\bar{\psi}(z)A^\rho(u)\right>\\
&+i\bar{\sigma}^{\mu\nu}\left<T\psi(y)\bar{\psi}(z)A^\rho(u)\right>\delta^4(x-y)+i\left<T\psi(y)\bar{\psi}(z)A^\rho(u)\right>\bar{\sigma}^{\mu\nu}\delta^4(x-z).
\end{aligned}\end{equation}

The contribution of $-2r\partial_\rho\partial^\rho F^{\mu\nu}$ is thus recast to be the modification of the coefficient of the curl of $N\left[j^\nu\right]$ and a new contact term, $-2ir\left<T\psi(y)\bar{\psi}(z)\right>\left(\partial^\mu_x g^{\nu\rho}-\partial^\nu_x g^{\mu\rho}\right)\delta^4(x-u)$.

We recall that contact terms come from equal time commutation relations of operators like $j^\mu$ and elementary fields $\psi,\bar{\psi},A^\rho$ in the canonical framework, and it is thus concluded that there is a non-canonical contribution to $\left[N\left[j^i\right]\left(\vec{x},t\right),A^\rho\left(\vec{y},t\right)\right]$ :
\begin{equation}\begin{aligned}
\delta\left(x^0-y^0\right)\left[N\left[j^i\right]\left(\vec{x},x^0\right),A^\rho\left(\vec{y},y^0\right)\right]=&\frac{-2i\tilde{r}}{1-2b-2g\tilde{r}}\left(\partial^0_x g^{i\rho}-\partial^i_x g^{0\rho}\right)\delta^4(x-y).
\end{aligned}\end{equation}
According to Schwinger\cite{schwinger}, this non-zero commutator is required so as to not conflict with the existence of a vacuum state. As definite operators, the commutator of $N\left[j^\nu\right]$ and $A^\rho$ should not have arbitrariness. Then, $\frac{-2i\tilde{r}}{1-2b-2g\tilde{r}}$ is fixed and thus $\tilde{r}$ is fixed. However, $\tilde{s}$ remains arbitrary.

Furthermore, we can also work out Schwinger terms for $\left[N\left[j^i\right],N\left[j^\rho\right]\right]$. We consider the following WTI \footnote{The last term is easily derived using Zimmermann-like identities in dimensional renormalization proposed by Bonneau\cite{bonneau1,bonneau2}. Note that there are no contact terms corresponding to $\psi,\bar{\psi},A^\sigma$ in this situation.}:
\begin{equation}\begin{aligned}
&\left(1-2b-2g\tilde{r}\right)\left(\partial^\mu_x\left<T N\left[j^\nu\right](x)N\left[j^\rho\right](y)\right>-\partial^\nu_x \left<TN\left[j^\mu\right](x)N\left[j^\rho\right](y)\right>\right)\\
=&2\left(1-c\right)\left<T\tilde{N}\left[\epsilon^{\mu\nu\rho\sigma}\bar{\psi}\gamma_\sigma\gamma_5 iD_\rho\psi\right](x)N\left[j^\rho\right](y)\right>+2m\left(1-f\right)\left<T\tilde{N}\left[\bar{\psi}\bar{\sigma}^{\mu\nu}\psi\right](x)N\left[j^\rho\right](y)\right>\\
&-2\tilde{s}m^2\left<TF^{\mu\nu}N\left[j^\rho\right](y)\right>\\
&-2\left(u\partial^\alpha_y\partial_\alpha^y+vm^2\right)\left(\partial^\mu_y g^{\nu\rho}-\partial^\nu_y g^{\mu\rho}\right)\delta^4(x-y),
\end{aligned}\label{tWTI-DR4}\end{equation}
where $u\equiv u'/(1-a),v\equiv v'/(1-a)$ and 
\begin{equation}\begin{aligned}
u'=&\frac{1}{-288i}\frac{\partial}{\partial q_\delta}\frac{\partial}{\partial q^\delta}\frac{\partial}{\partial q_\sigma}\mathrm{r.s.p.}\overline{\left<TN\left[\check{g}_{\alpha\beta}\left(\bar{\psi}\bar{\sigma}^{\mu\nu}\gamma^\alpha iD^\beta\psi\right)\right]N\left[\tilde{j}^\rho\right](q)\right>^{\mathrm{prop}}}|_{\check{g},q=0}\times\left(\bar{g}_{\mu\rho}\bar{g}_{\nu\sigma}-\bar{g}_{\mu\sigma}\bar{g}_{\nu\rho}\right)\\
=&r'/g,\\
v'=&\frac{1}{-12im^2}\frac{\partial}{\partial q_\sigma}\mathrm{r.s.p.}\overline{\left<TN\left[\check{g}_{\alpha\beta}\left(\bar{\psi}\bar{\sigma}^{\mu\nu}\gamma^\alpha iD^\beta\psi\right)\right]N\left[\tilde{j}^\rho\right](q)\right>^{\mathrm{prop}}}_{\check{g},q=0}\times\left(\bar{g}_{\mu\sigma}\bar{g}_{\nu\rho}-\bar{g}_{\mu\rho}\bar{g}_{\nu\sigma}\right)\\
=&s'/g.
\end{aligned}\end{equation}
Therefore, the Schwinger terms of $\left[N\left[j^i\right],N\left[j^\rho\right]\right]$ are
\begin{equation}\begin{aligned}
&\delta\left(x^0-y^0\right)\left[N\left[j^i\right]\left(\vec{x},x^0\right),N\left[j^\rho\right]\left(\vec{y},y^0\right)\right]\\
=&\left(\frac{-2ir/g}{1-2b-2g\tilde{r}}\partial^\alpha_y\partial^y_\alpha+\frac{-2is/g}{1-2b-2g\tilde{r}}m^2\right)\left(\partial^0_y g^{i\rho}-\partial^i_y g^{0\rho}\right)\delta^4(x-y).
\end{aligned}\end{equation}
Taking $\rho=0$, on the one-loop level, we have
\begin{equation}\begin{aligned}
&\delta\left(x^0-y^0\right)\left[N\left[j^0\right]\left(\vec{x},x^0\right),N\left[j^i\right]\left(\vec{y},y^0\right)\right]\\
=&-\left(\frac{i}{12\pi^2}\partial^\alpha_x\partial^x_\alpha+\frac{im^2}{2\pi^2}\right)\partial^i_x\delta^4(x-y).
\end{aligned}\end{equation}
This is comparable to results published in earlier papers. As an example, in \cite{ST1}, it was obtained that $\left<\left[j^0\left(\vec{x},0\right),j^i\left(0\right)\right]\right>_0$ is ((10) in \cite{ST1}):
\begin{equation}\begin{aligned}
\left<\left[j^0\left(\vec{x},0\right),j^i\left(0\right)\right]\right>_0=\infty\partial^i\delta\left(\vec{x}\right)+\frac{i}{12\pi^2}\partial^i\Delta\delta\left(\vec{x}\right),
\end{aligned}\label{eq:ST-a}\end{equation}
using spectral representation, where $\Delta\equiv\nabla^2$. However, we get a finite and covariant result, in contrast with the infinite and non-covariant result obtained in \cite{ST1}. In any event, the reproduction of the term\footnote{The Schwinger term (\ref{eq:ST-a}) was also obtained in \cite{ST2} using the BJL (Bjorken, Johnson and Low) method.} $\frac{i}{12\pi^2}\partial^i\Delta\delta\left(\vec{x}\right)$ implies that transverse anomalies are closely related to Schwinger terms.

\section{Comments on previous articles}\label{sec:comments}

There are papers \cite{kondo1997,he2001,sun2003} on the anomalies of the tWTI, but none found an anomaly for the vector tWTI. Additionally, \cite{he2007} examined the vector tWTI to one-loop order in dimensional regularization and concluded that there was no anomaly. In fact, \cite{he2007} has noted that $\epsilon^{\mu\nu\rho\sigma}\gamma_\rho\gamma_5$ should be replaced by $-\frac{1}{2}\left\{\gamma^\rho,\sigma^{\mu\nu}\right\}$ to ensure tWTI is still established; otherwise, on one-loop order, additional terms of the form a divergent integral multiplied by $(d-4)$ appear\footnote{However, \cite{he2007} did not look into this one-loop anomalous term.}. Equivalently, they adopted schemes that absorb all transverse anomalies into $N\left[\epsilon^{\mu\nu\rho\sigma}\bar{\psi}\gamma_\sigma\gamma_5 iD_\rho\psi\right]$; i.e.,  what appeared in the tWTI in their paper is not $N\left[\epsilon^{\mu\nu\rho\sigma}\bar{\psi}\gamma_\sigma\gamma_5 iD_\rho\psi\right]$ but $N\left[-i\bar{\psi}\left\{\gamma^\rho,\sigma^{\mu\nu}\right\}D_\rho\psi\right]$. However, the $(d-4)$ term they discovered is actually a spurious anomaly corresponding to corrections of coefficients of terms existing in the tWTI, such as $N\left[\epsilon^{\mu\nu\rho\sigma}\bar{\psi}\gamma_\sigma\gamma_5 iD_\rho\psi\right]$, because they only examined one-loop diagrams with two external fermion legs and did not note the crucial diagram with one external photon legs that generates transverse anomalies. We calculated this missing diagram and obtained exactly the results of $\zeta$ function regularization in App.\ref{app:zeta}.

In \cite{kondo1997}, the author identified the transverse vector transformation (\ref{eq:var-tWTI}) as the ``local Lorentz transformation''\footnote{In fact, only the spinor part.}, and there was thus no possibility of an anomaly in the vector tWTI due to Lorentz invariance. However, the Lorentz transformation of the Dirac fermion mismatches the transverse transformation (\ref{eq:var-tWTI}) in \emph{signs}. The spinor part of the Lorentz boost of the fermion is\cite{peskin} $\delta\psi(x)=+\frac{i}{4}\alpha(x)\epsilon_{\mu\nu}\sigma^{\mu\nu}\psi(x),\delta\bar{\psi}(x)=-\frac{i}{4}\alpha(x)\epsilon_{\mu\nu}\bar{\psi}(x)\sigma^{\mu\nu}$, where Jacobians of $\psi$ and $\bar{\psi}$ cancel each other out, regardless of whether $\mathrm{tr}[\sigma^{\mu\nu}]\delta^4(x-x)$ vanishes. According to our derivation, $\mathrm{tr}[\sigma^{\mu\nu}]\delta^4(x-x)$ is not zero, and thus transverse transformation, both signs of which are \emph{positive}, cannot be protected by Lorentz symmetry to be free of anomalies.

The point-splitting method was used in \cite{he2001}. A spurious transverse axial anomaly was proposed but corrected in \cite{sun2003}. Meanwhile, \cite{he2001} gave a expression for the ``vanishing'' transverse vector anomaly; however, following this formulation, we get a \emph{non-vanishing} result. Equation (12) of \cite{he2001} is\footnote{$U_P(x',x)=\mathrm{exp}\{-ig\int_x^{x'}dy\cdot A\}$ differs from ours in sign, because \cite{he2001} assigned $\overrightarrow{D}_\mu=\overrightarrow{\partial}_\mu+igA_\mu$.}
\begin{equation}\begin{aligned}
\partial^\mu j^\nu(x)-\partial^\nu j^\mu(x)=&\lim_{x'\rightarrow x}i\left(\partial^x_\lambda-\partial^{x'}_\lambda\right)\epsilon^{\lambda\mu\nu\rho}\bar{\psi}(x')\gamma_\rho\gamma_5U_P(x',x)\psi(x)\\
&+\mathrm{Symm}\lim_{\epsilon\rightarrow0}\left\{\bar{\psi}(x+\epsilon/2)\left[-ig\left(\gamma^\nu F^{\mu\rho}(x)-\gamma^\mu F^{\nu\rho}(x)\right)\epsilon_\rho\right]\psi(x-\epsilon/2)\right\}.
\end{aligned}\end{equation}
Using\cite{peskin} $\left<\psi(x)\bar{\psi}(y)\right>\propto \frac{\gamma^\sigma(x-y)_\sigma}{(x-y)^4}$,
and \cite{he2001} $\mathrm{Symm}\lim_{\epsilon\rightarrow0}\left\{\frac{\epsilon_\rho\epsilon_\sigma}{\epsilon^2}\right\}=\frac{1}{4}g^{\mu\nu}$, we finish the calculation of the last term :
\begin{equation}\begin{aligned}
&\mathrm{Symm}\lim_{\epsilon\rightarrow0}\left\{\bar{\psi}(x+\epsilon/2)\left[-ig\left(\gamma^\nu F^{\mu\rho}(x)-\gamma^\mu F^{\nu\rho}(x)\right)\epsilon_\rho\right]\psi(x-\epsilon/2)\right\}\\
\propto&\mathrm{Symm}\lim_{\epsilon\rightarrow0}\mathrm{tr}\left[\left(\gamma^\nu F^{\mu\rho}(x)-\gamma^\mu F^{\nu\rho}(x)\right)\gamma^\sigma\right]\frac{\epsilon_\sigma\epsilon_\rho}{\epsilon^4}\\
\propto&\lim_{\epsilon\rightarrow0}\left(g^{\nu\sigma}F^{\mu\rho}(x)-g^{\mu\sigma}F^{\nu\rho}(x)\right)g_{\sigma\rho}\frac{1}{\epsilon^2}\\
\propto&\lim_{\epsilon\rightarrow0}\frac{1}{\epsilon^2}F^{\mu\nu}(x)\neq0.
\end{aligned}\label{eq:non-vanishing}\end{equation}
Moreover, because the above result is quadratically divergent, we need to expand $F^{\mu\nu}(x\pm\epsilon/2)$ in intermediate steps (see App.\ref{app:ps}) to $\mathcal{O}(\epsilon^4)$ to extract finite contributions, which means (\ref{eq:non-vanishing}) is incomplete\footnote{However, even if we go to $\mathcal{O}(\epsilon^4)$, arbitrariness of the coefficient of $\partial^\rho\partial_\rho F^{\mu\nu}$ that originates from the arbitrariness of $a\in\mathbbm{R}$ in $\bar{\psi}(x+(a+1)\epsilon)\gamma^\mu\psi(x+a\epsilon)$ prevents the point-splitting method from working for transverse anomalies; see App.\ref{app:ps}.}.

In brief, \cite{he2001} partially worked out transverse anomalies. It is a pity that the non-vanishing result (\ref{eq:non-vanishing}) was omitted in \cite{he2001}.

Pauli--Villars regularization\cite{pauli-villars} was applied to calculating transverse anomalies in \cite{sun2003}. Unfortunately \cite{sun2003} forgot a vital procedure in Pauli--Villars regularization and thus missed transverse anomalies. This step expresses  any amplitude with  its regularized form so that anomalies may appear from the WTI with mass terms\cite{pauli-villars,bertlmann}, which is the case for the vector tWTI (\ref{tWTI-BG2}). We consider any WTI with the form
\begin{equation}
A=mB+C,
\end{equation}
where $m$ is some particle's mass. Pauli--Villars regularization requires\cite{bertlmann} regularized WTI to be made up of regularized amplitudes
\begin{equation}
f^{\mathrm{phys}}=\lim_{M\rightarrow\infty}f_m-f_M,
\end{equation}
where $f$ denotes any amplitude while $f_m,f_M$ respectively denote amplitudes calculated with physical mass $m$ and regulated mass $M$. Then, if we proved the bare WTI
\begin{equation}
A_m-A_M=mB_m+C_m-MB_M-C_M,
\end{equation}
the regularized WTI may acquire an anomaly
\begin{equation}\begin{aligned}
&A^{\mathrm{phys}}=A_m-A_M=mB_m+C_m-MB_M-C_M\\
=&mB^{\mathrm{phys}}+C^{\mathrm{phys}}+(m-M)B_M.
\end{aligned}\end{equation}

Indeed, on the basis of the proof of the bare tWTI in \cite{sun2003}, we worked out transverse anomalies that \cite{sun2003} ignored, see App.\ref{app:pv}.

\section{Comparison between Symmetry Anomalies and Non-symmetry Anomalies}\label{sec:comparison}

In fact, $\partial^\mu j^\nu-\partial^\nu j^\mu$ on the left of tWTI (\ref{tWTI-DR2}) can be recast into the divergence of some current :
\begin{equation}
\partial^\mu j^\nu-\partial^\nu j^\mu=-\frac{i}{2}\partial_\rho\left(\bar{\psi}\left[\gamma^\rho,\sigma^{\mu\nu}\right]\psi\right).
\end{equation}

In addition to transverse anomalies and the mass term, factors that prevent $\bar{\psi}\left[\gamma^\rho,\sigma^{\mu\nu}\right]\psi$ from being a conserved current include another four-dimensional operator; i.e., $N\left[\epsilon^{\mu\nu\rho\sigma}\bar{\psi}\gamma_\sigma\gamma_5 iD_\rho\psi\right]$. Without this operator, the current will become an anomalous partial conserved current, which is the case for the tWTI in two-dimensional QED\footnote{In two-dimensional QED with a massless fermion, $\left\{\gamma^\rho,\sigma^{\mu\nu}\right\}=0$ and $j^\mu_5=-\epsilon^{\mu\nu}j_\nu$ owing to $\sigma^{\mu\nu}=i\epsilon^{\mu\nu}\gamma_5$ and $\gamma^\mu\gamma_5=-\epsilon^{\mu\nu}\gamma_\nu$, and thus $-\frac{i}{2}\partial_\rho\left(\psi\left[\gamma^\rho,\sigma^{\mu\nu}\right]\psi\right)=\partial^\mu j^\nu-\partial^\nu j^\mu=\epsilon^{\mu\nu}\partial_\rho j^\rho_5=-\frac{g}{2\pi}\epsilon^{\mu\nu}\epsilon_{\rho\sigma}F^{\rho\sigma}=\frac{g}{\pi}F^{\mu\nu}$\cite{peskin}. Therefore, both $j^\mu_5$ and $\partial_\rho\left(\psi\left[\gamma^\rho,\sigma^{\mu\nu}\right]\psi\right)$ are anomalous partial conserved currents, and as a consequence, the tWTI in two-dimensional QED is \emph{not} a non-symmetry WTI.}. Therefore, the essential difference between a symmetry WTI and a non-symmetry WTI is the presence of extra four-dimensional operators, besides operators on the left of the WTI\footnote{These operators are usually in the form of a derivative of some three-dimensional operator as is the case of chiral ($\partial_\mu j^\mu_5$) and trace ($\partial_\mu\left(x_\nu\theta^{\mu\nu}\right)$) anomalies.}and anomalies.

It is exactly the extra four-dimensional operators that render the arbitrariness of transverse anomalies. Obviously, this makes sense also for any other non-symmetry anomalies that have extra four-dimensional operators.

We may proceed further. It is also possible to change coefficients of chiral or trace anomalies at will, as long as we absorb $N\left[\tilde{F}^{\mu\nu}A_\nu\right]$ or $N\left[\bar{g}^{\mu\nu}F^{\rho\sigma}F_{\rho\sigma}\right]$ into $N\left[\bar{\psi}\gamma^\mu\gamma_5\psi\right]$ or $N\left[\theta^{\mu\nu}\right]$ without considering gauge invariance or energy conservation. However, it is just these non-trivial properties or symmetries satisfied by $N\left[\bar{\psi}\gamma^\mu\gamma_5\psi\right]$ and $N\left[\theta^{\mu\nu}\right]$ that prevent other operators such as anomaly terms to be absorbed into them, thus \emph{protected} chiral and trace anomalies such that their coefficients cannot be adjusted arbitrarily.  Therefore, once we find some physical meanings or symmetries for $N\left[\epsilon^{\mu\nu\rho\sigma}\bar{\psi}\gamma_\sigma\gamma_5 iD_\rho\psi\right]$, coefficients of transverse anomalies may be fixed naturally. As shown in Sec.\ref{sec:ST}, coefficients of transverse anomalies may be fixed partially by resorting to Schwinger terms, but more general results for remaining $\tilde{s}$ and other non-symmetry anomalies require deeper research.

\section{Conclusion and discussion}\label{sec:conclusion}

We discuss the extension of anomalies to cover those in WTIs that are not formed by symmetry transformations, beyond the scope of symmetry, taking the explicit example of anomalies in tWTIs. Both background field (one-loop) analyses in Sec.\ref{sec:derivation-fujikawa} (together with some one-loop calculation in App.\ref{app:pv} and App.\ref{app:ps}) and renormalization to all orders in dimensional renormalization in Sec.\ref{sec:derivation-DR} indicate the existence of transverse anomalies, and we locate where anomalies hid when \cite{kondo1997,he2001,sun2003,he2007} stated the non-existence of the transverse anomalies in the vector tWTI on one-loop order. The scheme dependence of coefficients of transverse anomalies is also concluded temporarily, and this is partially solved by considering Schwinger terms as in Sec.\ref{sec:ST}. This needs to be investigated further.

So far, the anomaly in all types of the local linear transformation of fermion fields\footnote{The local and linear transformation of $\psi(x),\bar{\psi}(x)$ must be $\delta\psi(x)=\alpha(x)\Omega\psi(x),\delta\bar{\psi}(x)=\alpha(x)\bar{\psi}(x)\widetilde{\Omega}$, where $\Omega$ and $\widetilde{\Omega}$ are a linear combination of $\gamma$ matrices and hence of $\mathbbm{1},\gamma^\mu,[\gamma^\mu,\gamma^\nu],\gamma^\mu\gamma_5,\gamma_5$. However, the traces of odd number $\gamma$ matrices are zero; i.e., $\mathrm{tr}[\gamma^\mu]\delta^4(x-x)$ and $\mathrm{tr}[\gamma^\mu,\gamma_5]\delta^4(x-x)$ are zero even after regularization.} (not all symmetry transformations) has been exhausted. There are only three non-zero anomalies; see Table.\ref{tab:1} (in Fujikawa's style for simplicity).
\begin{table}[ht]
\centering
\caption{Non-trivial anomalies in all types of local linear transformation of fermion fields.}
\label{tab:1}
\begin{tabular}{|c|c|c|}
\hline
\mbox{anomaly type}&\mbox{``bare" expression}&\mbox{in 4-dim $SU(N)$ QCD (1 loop results)}\\
\hline
\mbox{trace anomaly\cite{fujikawa1980,fujikawa2004}}&$\mathrm{tr}[\mathbbm{1}]\delta^4(x-x)$&$\frac{g^2}{48\pi^2}F_a^{\mu\nu}F_{a\mu\nu}$\\
\hline
\mbox{transverse anomalies}&$\mathrm{tr}\left\{t_a[\gamma^\mu,\gamma^\nu]\right\}\delta^4(x-x)$&$-\frac{gm^2}{4\pi^2}F_a^{\mu\nu}-\frac{g}{24\pi^2}\mathcal{D}^\rho\mathcal{D}_\rho F_a^{\mu\nu}-\frac{g^2}{8\pi^2}C_{abc}F_b^{\mu\rho}F_{c\rho}^{\ \ \nu}$\\
\hline
\mbox{chiral anomaly\cite{fujikawa1979,fujikawa2004}}&$\mathrm{tr}[t_a\gamma_5]\delta^4(x-x)$&$\frac{g^2}{32\pi^2}\epsilon_{\mu\nu\rho\sigma}F_b^{\mu\nu}F_c^{\rho\sigma}\mathrm{tr}[t_at_bt_c]$\\
\hline
\end{tabular}
\end{table}

Table.\ref{tab:1} shows that the transverse anomalies have many more types of operators than the trace anomaly and chiral anomaly. In particular, the $C_{abc}F_b^{\mu\rho}F_{c\rho}^{\ \ \nu}$ term may have some effect on the present scheme\cite{qin2013,qin2014,albino2019} making use of tWTI. However, in this scheme, the other two terms in non-Abelian transverse anomalies and the whole Abelian transverse anomalies (where $C_{abc}=0$) have no places to plug in, because the general method\cite{qin2013,qin2014} is to contract $\epsilon_{\alpha\mu\nu\beta}t_\alpha q_\beta,\epsilon_{\alpha\mu\nu\beta}\gamma_\alpha q_\beta$ to the vector tWTI\footnote{So far, \cite{qin2013,qin2014} discussed only the Abelian case. And here we use the Abelian tWTI for an explanation.} in the momentum space\footnote{Eq.(4) in\cite{qin2013}, in the Euclidean metric; $q\equiv k-p,t\equiv k+p$.} :
\begin{equation}\begin{aligned}
q_\mu&\Gamma_\nu(k,p)-q_\nu\Gamma_\mu(k,p)=S^{-1}(p)\sigma_{\mu\nu}+\sigma_{\mu\nu}S^{-1}(k)\\
&+2im\Gamma_{\mu\nu}(k,p)+t_\lambda\epsilon_{\lambda\mu\nu\rho}\Gamma^A_{\rho}(k,p)+A^V_{\mu\nu}(k,p),
\end{aligned}\end{equation}
such that the identically vanishing left-hand side and the contracted right-hand side serve as constraints for axial vertex $\Gamma^A_{\rho}(k,p)$ to be solved. Therefore, additional terms\footnote{$\widetilde{A}_\mu(k,p)$ is $\left<TA_\mu(x)\psi(y)\bar{\psi}(z)\right>$ in momentum space.} $q^2q_{[\mu}\widetilde{A}_{\nu]}(k,p)$ and $q_{[\mu}\widetilde{A}_{\nu]}(k,p)$ of the Abelian transverse anomalies all vanish after contraction with $\epsilon_{\alpha\mu\nu\beta}q_\alpha$ because $\epsilon_{\alpha\mu\nu\beta}q_\alpha q_\mu=0$. The Abelian transverse anomalies are thus neglectable in current schemes\cite{qin2013,qin2014} making use of tWTI. However, even if the ordinary derivative parts\footnote{Of course, the gauge field parts (where no ordinary derivative appears) of $F_{a}^{\mu\nu}$ and $\mathcal{D}^\rho\mathcal{D}_{\rho}F_a^{\mu\nu}$ are not zero in general, but these are  not gauge covariant and thus may be zero by proper choice of gauge. However, $C_{abc}F_b^{\mu\rho}F_{c\rho}^{\ \ \nu}$ is gauge covariant and its contribution cannot be neglected.} of $F_{a}^{\mu\nu}$ and $\mathcal{D}^\rho\mathcal{D}_{\rho}F_a^{\mu\nu}$ vanish owing to the same reason as the case of the Abelian tWTI, the non-Abelian transverse anomalies have a non-vanishing contribution from $C_{abc}F_b^{\mu\rho}F_{c\rho}^{\ \ \nu}$ in this scheme because $C_{abc}F_b^{\mu\rho}F_{c\rho}^{\ \ \nu}$ is not of the form $q_{[\mu} f_{\nu]}(k,p)$ where $f_\nu(k,p)$ is some operator's Fourier transformed Green's function. Unfortunately, the Abelian approximation (i.e., $\Gamma^\mu_a(\mbox{non-Abelian})\approx t_a\Gamma^\mu(\mbox{Abelian})$) remains the backbone\cite{qin2014,qin2014,liu2017,liu2019}. However, once we begin to attack the non-Abelian quark-gluon vertex directly using the non-abelian tWTI (\ref{eq:non-abelian}), the transverse anomaly will take some responsibility. Further more, other possible applications to the transverse anomaly are being researched.

\acknowledgments

The work of Q. Wang was supported in part by the National Key Research and Development Program of China (Grant No.2017YFA0402200) and the National Natural Science Foundation of China (Grant No. 11475092).

\bibliography{ref}

\appendix

\section{Non-Abelian Generalization and one-loop Calculation through $\zeta$ function regularization}\label{app:zeta}

The generalization of transverse anomalies to the non-Abelian case (with gauge group $SU(N)$) is straightforward. Within Fujikawa's framework, using the Lagrangian\cite{peskin} :
\begin{equation}\begin{aligned}
\mathscr{L}_{SU(N)}=&\bar{\psi}\left(i\slashed{D}-m_0\right)\psi,\ D_\mu\equiv\frac{1}{2}\left(\overrightarrow{\partial}_\mu-\overleftarrow{\partial}_\mu\right)-igt_aA_{a\mu},
\end{aligned}\end{equation}
applying variations of fermion fields :
\begin{equation}
\delta\psi(x)=\frac{1}{4}\epsilon_{\mu\nu}\alpha_a(x)t_a\sigma^{\mu\nu}\psi(x),\ \delta\bar{\psi}(x)=\frac{1}{4}\epsilon_{\mu\nu}\alpha_a(x)\bar{\psi}(x)t_a\sigma^{\mu\nu},
\label{eq:var-sun}\end{equation}
and considering the transformation Jacobian (of which we present a concrete calculation later), we get :
\begin{equation}\begin{aligned}
&\mathcal{D}^\mu\left<j^\nu_a(x)\right>_A-\mathcal{D}^\nu\left<j^\mu_a(x)\right>_A\\
=&\left<\epsilon^{\mu\nu\rho\sigma} \bar{\psi}(x)\gamma_\sigma\gamma_5\left\{iD_\rho,t_a\right\}\psi(x)\right>_A+2m_0\left<\bar{\psi}(x)\sigma^{\mu\nu}t_a\psi(x)\right>_A\\
&-\frac{g_0m_0^2}{4\pi^2}F^{\mu\nu}_a(x)-\frac{g_0}{24\pi^2}\mathcal{D}^\rho\mathcal{D}_\rho F_a^{\mu\nu}(x)-\frac{g^2_0}{8\pi^2}C_{abc}F_b^{\mu\rho}(x)F_{c\rho}^{\ \ \nu}(x).
\end{aligned}\label{eq:non-abelian}\end{equation}
Renormalization of the above non-Abelian tWTI is left as further work.

We next calculate the one-loop transverse anomalies, $-2i\mathrm{tr}[t_a\sigma^{\mu\nu}]\delta^4(x-x)$, through $\zeta$ function regularization. Identification of the transformation Jacobian of (\ref{eq:var-sun}) and (\ref{eq:var-tWTI}) to be  $-2i\mathrm{tr}[t_a\sigma^{\mu\nu}]\delta^4(x-x)$ and $-2i\mathrm{tr}[\sigma^{\mu\nu}]\delta^4(x-x)$ is straightforward after using the following equation (recall that $\ln(1+x)\cong x$ when $x\ll 1$):
\begin{equation}\begin{aligned}
&\mathrm{Det}\left[\frac{\delta}{\delta\psi(y)}\left(\psi(x)+\frac{1}{4}\epsilon_{\mu\nu}\alpha_a(x)t_a\sigma^{\mu\nu}\psi(x)\right)\right]\\
=&\mathrm{Det}\left[\delta^4(x-y)+\frac{1}{4}\epsilon_{\mu\nu}\alpha_a(x)t_a\sigma^{\mu\nu}\delta^4(x-y)\right]\\
=&\mathrm{exp}\left\{\mathrm{tr}\ln\left[\delta^4(x-y)+\frac{1}{4}\epsilon_{\mu\nu}\alpha_a(x)t_a\sigma^{\mu\nu}\delta^4(x-y)\right]\right\}\\
\cong&\mathrm{exp}\left\{\frac{1}{4}\epsilon_{\mu\nu}\alpha_a(x)\mathrm{tr}\left[t_a\sigma^{\mu\nu}\right]\delta^4(x-x)\right\}.
\end{aligned}\label{eq:jacobian}\end{equation}

$\sigma^{\mu\nu}\equiv\frac{i}{2}[\gamma^\mu,\gamma^\nu]$, and it is thus enough to calculate $\mathrm{tr}\left\{t_a[\gamma^\mu,\gamma^\nu]\right\}\delta^4(x-x)$. The combination of Fujikawa's approach\cite{fujikawa1979} and $\zeta$ function regularization\cite{bertlmann,zeta} leads to :
\begin{equation}\begin{aligned}
&\left\{\mathrm{tr}\left\{ t_a[\gamma^\mu,\gamma^\nu]\right\}\delta^4(x-x)\right\}_\zeta\\
=&\frac{d}{ds}\Bigg|_{s=0}\frac{s}{\Gamma(s)}\int_0^\infty d\tau\ \tau^{s-1}\left\{\mathrm{tr}\left[e^{-\left(\slashed{D}_x^2+m^2\right)\tau}t_a[\gamma^\mu,\gamma^\nu]\right]\int\frac{d^4k}{(2\pi)^4}e^{-ik\cdot(x-y)}\right\}_{y\rightarrow x}\\
=&\frac{d}{ds}\Bigg|_{s=0}\frac{s}{\Gamma(s)}\int_0^\infty d\tau\ \tau^{s-1}e^{-m^2\tau}\frac{1}{(\sqrt{\tau})^4}\int\frac{d^4k}{(2\pi)^4}\\
&\times\mathrm{tr}\left[t_a[\gamma^\mu,\gamma^\nu]\mathrm{exp}\left(k^2+2ik\cdot D\sqrt{\tau}-D^2\tau+\frac{i}{4}gt_b[\gamma^\rho,\gamma^\sigma]F_{b\rho\sigma}\tau\right)\right].
\end{aligned}\label{eq:zeta1}\end{equation}
In the last step, we used $\slashed{D}^2=D^2-\frac{1}{4}igt_a[\gamma^\mu,\gamma^\nu]F_{a\mu\nu}$ and rescaled $k\rightarrow k/\sqrt{\tau}$.

We note that\footnote{In (\ref{eq:zeta1}), we strip $\frac{d}{ds}|_{s=0}\frac{s}{\Gamma(s)}\int_0^\infty d\tau\ \tau^{s-1}e^{-m^2\tau}$ away and substitute $\sqrt{\tau}$ with $1/\Lambda$, thus arriving at the original Fujikawa's method. From (\ref{eq:zeta_reg}), it is obvious that $\Lambda^2$ in the final results, like (\ref{eq:fujikawa}), is effectively regularized to be $-m^2$ through $\zeta$ function regularization.}
\begin{equation}\begin{aligned}
&\frac{d}{ds}\Bigg|_{s=0}\frac{s}{\Gamma(s)}\int_0^\infty d\tau\ \tau^{s-1}e^{-m^2\tau}\frac{1}{(\sqrt{\tau})^4}(\sqrt{\tau})^n\\
=&\frac{d}{ds}\Bigg|_{s=0}\frac{s\Gamma(s+n/2-2)}{\Gamma(s)}(m^2)^{2-s-\frac{n}{2}}\\
=&\left\{\begin{matrix}
\frac{m^4}{2},&n=0;\\
-m^2,&n=2;\\
1,&n=4;\\
0,&\mbox{otherwise.}
\end{matrix}\right.
\end{aligned}\label{eq:zeta_reg}\end{equation}
Therefore, the only contributing terms in (\ref{eq:zeta1}) are those proportional to $(\sqrt{\tau})^0,(\sqrt{\tau})^2,(\sqrt{\tau})^4$ in expansion of the exponential inside the trace.

The $(\sqrt{\tau})^0$ term is zero because $\mathrm{tr}\left\{[\gamma^\mu,\gamma^\nu]\right\}=0$. The $(\sqrt{\tau})^2$ term is (after finishing $\frac{d}{ds}|_{s=0}$)
\begin{equation}\begin{aligned}
(-m^2)\mathrm{tr}[t_at_b]\mathrm{tr}\left\{[\gamma^\mu,\gamma^\nu][\gamma^\rho,\gamma^\sigma]\right\}\int\frac{d^4k}{(2\pi)^4}e^{k^2}\frac{i}{4}gF_{b\rho\sigma}.
\end{aligned}\label{eq:tau2}\end{equation}

The $(\sqrt{\tau})^4$ term is
\begin{equation}\begin{aligned}
&\mathrm{tr}[t_at_bt_c]\mathrm{tr}\left\{[\gamma^\mu,\gamma^\nu][\gamma^\rho,\gamma^\sigma][\gamma^\alpha,\gamma^\beta]\right\}\int\frac{d^4k}{(2\pi)^4}e^{k^2}\frac{1}{2!}\left(\frac{i}{4}g\right)^2F_{b\rho\sigma}F_{c\alpha\beta}\\
+&\mathrm{tr}\left\{[\gamma^\mu,\gamma^\nu][\gamma^\rho,\gamma^\sigma]\right\}\int\frac{d^4k}{(2\pi)^4}e^{k^2}\frac{1}{2!}\frac{i}{4}g(-1)\mathrm{tr}\left[t_a\left(D^2t_bF_{b\rho\sigma}+t_bF_{b\rho\sigma}D^2\right)\right]\\
+&\mathrm{tr}\left\{[\gamma^\mu,\gamma^\nu][\gamma^\rho,\gamma^\sigma]\right\}\int\frac{d^4k}{(2\pi)^4}e^{k^2}\frac{1}{3!}\frac{i}{4}g(2i)^2\\
&\times\mathrm{tr}\left[t_a\left((k\cdot D)^2t_bF_{b\rho\sigma}+(k\cdot D) t_b F_{b\rho\sigma}(k\cdot D)+t_bF_{b\rho\sigma}(k\cdot D)^2\right)\right].
\end{aligned}\label{eq:tau4}\end{equation}

We arrive at the final result by completing the integral and working out the trace :
\begin{equation}\begin{aligned}
&\left\{\mathrm{tr}\{t_a[\gamma^\mu,\gamma^\nu]\}\delta^4(x-x)\right\}_\zeta\\
=&-\frac{gm^2}{4\pi^2}F_a^{\mu\nu}-\frac{g}{24\pi^2}\mathcal{D}_\rho\mathcal{D}^\rho F_a^{\mu\nu}-\frac{g^2}{8\pi^2}C_{abc}F_{b}^{\mu\rho}F_{c\rho}^{\ \ \nu}.
\end{aligned}\end{equation}

The only difference between the Abelian case and non-Abelian case is the use of $\mathrm{tr}[t_at_b]=\frac{1}{2}\delta_{ab}$, which is not needed in the Abelian case.  The Abelian result is therefore \footnote{Additionally, note that $\mathrm{tr}[t_b]=0$ for $SU(N)$, such that there is no contribution in $\mathrm{tr}\ [\gamma^\mu,\gamma^\nu]\delta^4(x-x)$ from non-Abelian fields through observation on (\ref{eq:tau2})(\ref{eq:tau4}) with $t_a$ stripped away, using $\mathrm{tr}[t_bt_c]\mathrm{tr}\left\{[\gamma^\mu,\gamma^\nu][\gamma^\rho,\gamma^\sigma][\gamma^\alpha,\gamma^\beta]\right\}F_{b\rho\sigma}F_{c\alpha\beta}=0$.}
\begin{equation}\begin{aligned}
&\left\{\mathrm{tr}\ [\gamma^\mu,\gamma^\nu]\delta^4(x-x)\right\}_\zeta\\
=&-\frac{gm^2}{2\pi^2}F^{\mu\nu}-\frac{g}{12\pi^2}\partial_\rho\partial^\rho F^{\mu\nu},
\end{aligned}\label{eq:abelian anomaly}\end{equation}
where $F^{\mu\nu}\equiv\partial^\mu A^\nu-\partial^\nu A^\mu$ is the $U(1)$ gauge field.

\section{One-loop Calculations : Pauli--Villars Regularization}\label{app:pv}

This appendix calculates missing transverse anomalies in \cite{sun2003} adopting Pauli--Villars regularization, which is also the method used by \cite{sun2003}. As in \cite{sun2003}, we work with an external field $A_\mu$, and the Lagrangian is then again (\ref{eq:lag-BG}).

To verify (\ref{tWTI-BG2}), we go to momentum space and define
\begin{equation}\begin{aligned}
\Gamma^\mu(q)\equiv&\int d^4x\ e^{-iq\cdot x}\left<j^\mu(x)\right>_A,\\
N^{\mu\nu}(q)\equiv&\int d^4x\ e^{-iq\cdot x} 2\left<\bar{\psi}(x)\epsilon^{\mu\nu\rho\sigma}\gamma_\sigma\gamma_5 iD_\rho\psi(x)\right>_A,\\
T^{\mu\nu}(q)\equiv&\int d^4x\ e^{-iq\cdot x}\left<\bar{\psi}(x)\sigma^{\mu\nu}\psi(x)\right>.
\end{aligned}\end{equation}
The following ``bare" tWTI has been verified in \cite{sun2003} (for simplicity, we denote $m_0$ by $m$):
\begin{equation}\begin{aligned}
&iq^\mu\left(\Gamma^\nu_m+\Gamma^\nu_{M_1}-2\Gamma^\nu_{M_2}\right)-iq^\nu\left(\Gamma^\mu_m+\Gamma^\mu_{M_1}-2\Gamma^\mu_{M_2}\right)\\
=&N^{\mu\nu}_{m}+N^{\mu\nu}_{M_1}-2N^{\mu\nu}_{M_2}+2\left(mT^{\mu\nu}_{m}+M_1T^{\mu\nu}_{M_1}-2M_2T^{\mu\nu}_{M_2}\right).
\end{aligned}\end{equation}
Here, $\Gamma^\mu_m$ indicates an amplitude is calculated with fermion mass $m$; There are two subtractions because the leading divergence is quadratic. $M_1=\sqrt{m^2+2\Lambda^2},M_2=\sqrt{m^2+\Lambda^2},$ where $\Lambda$ serves as an effective cut-off, as in \cite{sun2003}.

However, in the spirit of Pauli--Villars regularization\cite{pauli-villars,bertlmann}, the WTI should be expressed by a regularized ``physical" amplitude for which any amplitude $f$ is defined as:
\begin{equation}
f_{\mathrm{phys}}\equiv\lim_{\Lambda\rightarrow\infty} f_m+rf_{M_1}-2sf_{M_2}.
\end{equation}
Here, $r$ and $s$ should be chosen to cancel out all divergences in $f$. For $T^{\mu\nu}$, it is easily seen that $r=m/M_1,s=m/M_2$ through direct analysis of the diagram on the lowest order of $A_\mu(x)$.)

The ``bare" identity then gets an extra term after assembling each amplitude into its regularized form :
\begin{equation}\begin{aligned}
&iq^\mu\Gamma^\nu_{\mathrm{phys}}-iq^\nu\Gamma^\mu_{\mathrm{phys}}\\
=&iq^\mu\left(\Gamma^\nu_m+\Gamma^\nu_{M_1}-2\Gamma^\nu_{M_2}\right)-iq^\nu\left(\Gamma^\mu_m+\Gamma^\mu_{M_1}-2\Gamma^\mu_{M_2}\right)\\
=&N^{\mu\nu}_{m}+N^{\mu\nu}_{M_1}-2N^{\mu\nu}_{M_2}+2m\left(T^{\mu\nu}_{m}+\frac{m}{M_1}T^{\mu\nu}_{M_1}-\frac{2m}{M_2}T^{\mu\nu}_{M_2}\right)\\
&+\left[2\left(M_1-\frac{m^2}{M_1}\right)T^{\mu\nu}_{M_1}-4\left(M_2-\frac{m^2}{M_2}\right)T^{\mu\nu}_{M_2}\right]\\
\equiv&N^{\mu\nu}_{\mathrm{phys}}-2mT^{\mu\nu}_{\mathrm{phys}}+\mathscr{A}^{\mu\nu}.
\end{aligned}\end{equation}

We will show that
\begin{equation}
\mathscr{A}^{\mu\nu}\equiv2\left(M_1-\frac{m^2}{M_1}\right)T^{\mu\nu}_{M_1}-4\left(M_2-\frac{m^2}{M_2}\right)T^{\mu\nu}_{M_2}
\end{equation}
is exactly the anomaly we obtained in App.\ref{app:zeta} up to quadratic divergences.

$T^{\mu\nu}$ is represented by the Feynman graphs\footnote{Recall that $C$ parity of $\bar{\psi}\sigma^{\mu\nu}\psi$ is odd, such that there are only $A^{2n+1}$ terms.} :
\begin{equation}
\includegraphics[width=6.5cm]{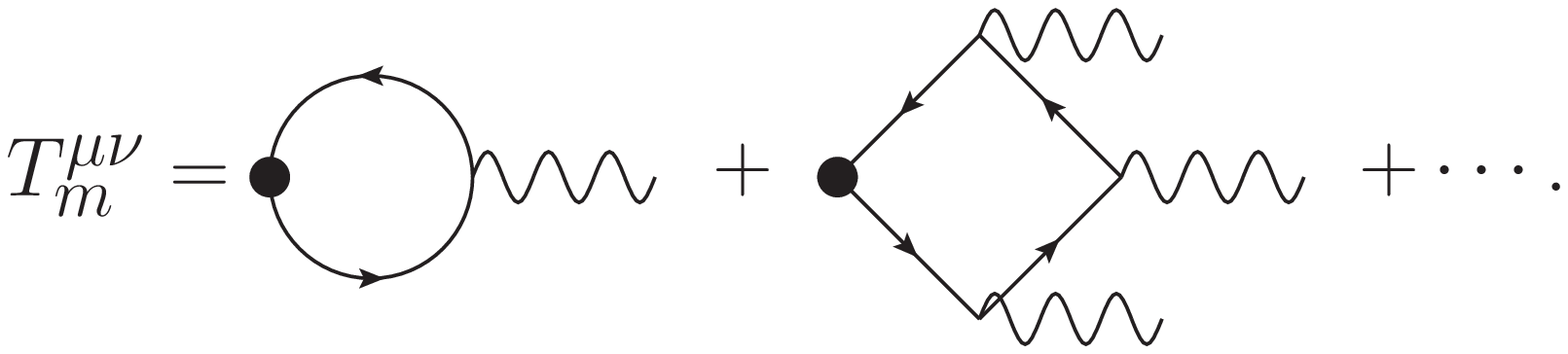}
\label{eq:diag-PV}\end{equation}
Gauge invariance in external photon legs and dimensional analysis tell us the diagram in (\ref{eq:diag-PV}) with $n$ photon legs diverges at worst like $Q^n A^{n} \Lambda^{3-2n}$ when $\Lambda\rightarrow\infty$, where $Q$ is some typical scale of external momenta and $A$ an abbreviation of $A^\mu(x)$. Therefore, the only term contributing to transverse anomalies is the smallest diagram :
\begin{equation}\begin{aligned}
&\mathscr{A}^{\mu\nu}=2\left(M_1-\frac{m^2}{M_1}\right)T^{\mu\nu}_{M_1,\mbox{the smallest diagram}}-4\left(M_2-\frac{m^2}{M_2}\right)T^{\mu\nu}_{M_2,\mbox{the smallest diagram}}\\
=&2\left(M_1-\frac{m^2}{M_1}\right)ig_0\int\frac{d^4p}{(2\pi)^4}\frac{\mathrm{tr}\left[\sigma^{\mu\nu}(\slashed{p}-\slashed{q}+M_1)\widetilde{\slashed{A}}(q)(\slashed{p}+M_1)\right]}{[(p-q)^2-m^2-2\Lambda^2][p^2-m^2-2\Lambda^2]}\\
&-4\left(M_2-\frac{m^2}{M_2}\right)ig_0\int\frac{d^4p}{(2\pi)^4}\frac{\mathrm{tr}\left[\sigma^{\mu\nu}(\slashed{p}-\slashed{q}+M_2)\widetilde{\slashed{A}}(q)(\slashed{p}+M_2)\right]}{[(p-q)^2-m^2-\Lambda^2][p^2-m^2-\Lambda^2]}\\
=&\left(iq^\mu \widetilde{A}^\nu-iq^\nu \widetilde{A}^\mu\right)\int_0^1dx\int_0^xdy\int_0^ydz\\
&\times g_0\frac{\Lambda^4}{\pi^2}\left(\frac{4}{[m^2+(1+x-y+z)\Lambda^2-y(1-y)q^2]}+\frac{2m^2+3\Lambda^2-[y^2+(1-y)^2]q^2}{[m^2+(1+x-y+z)\Lambda^2-y(1-y)q^2]^2}\right).
\end{aligned}\end{equation}

Taking the limit $\Lambda\rightarrow\infty$, we get
\begin{equation}
\mathscr{A}^{\mu\nu}=\left(\frac{g_0\ln2}{\pi^2}\Lambda^2-\frac{g_0}{2\pi^2}m^2+\frac{g_0}{12\pi^2}q^2\right)\left(iq^\mu \widetilde{A}^\nu-iq^\nu \widetilde{A}^\mu\right),\ \widetilde{A}^\mu(q)\equiv\int d^4x\ e^{-iq\cdot x}A^\mu(x).
\end{equation}

This is what we get in App.\ref{app:zeta} in momentum space, up to a quadratic divergent term. However, it is well-known\cite{pauli-villars,schwartz} that quadratic divergence in $\left<j^\mu\right>_A$ in the Pauli--Villars scheme corresponds to  an infinite photon mass, which must be subtracted to ensure gauge invariance. As long as we subtract quadratic divergence on both sides of\footnote{(\ref{eq:phys-PV}) holds at any $\Lambda$ so that quadratic divergences on the two sides are equal.}
\begin{equation}
iq^\mu\Gamma^\nu_{\mathrm{phys}}-iq^\nu\Gamma^\mu_{\mathrm{phys}}=N^{\mu\nu}_{\mathrm{phys}}-2mT^{\mu\nu}_{\mathrm{phys}}+\mathscr{A}^{\mu\nu},
\label{eq:phys-PV}\end{equation}
nothing is affected by quadratic divergence. Nevertheless, it is somehow confusing that there is no logarithmic divergence associated with this quadratic divergence. It is not easy (as far as we are concerned) to give a thorough explanation, but the situation may be summarized phenomenologically as an absence of logarithmic divergence is simply a signal of an anomaly because an anomaly is a local operator and leads to only polynomials of external momenta on one-loop order. (Note that the coefficient of an anomaly term is one-loop, and the matrix elements of the anomaly term are thus tree level.)

\section{One-loop Calculations : Point-splitting Method}\label{app:ps}

Similar to the case of chiral anomaly, the point-splitting method\cite{peskin} gives results for transverse anomalies. However, the dependence on the splitting ratio prevents this method from working for transverse anomalies.

The point-splitting method selects a special regularization for $j^\mu$ (where $a$ is a real number):
\begin{equation}
j^\mu(x)\rightarrow\bar{\psi}(x+(a+1)\epsilon)\gamma^\mu e^{ig\int_{x+a\epsilon}^{x+(a+1)\epsilon}dy\cdot A(y)}\psi(x+a\epsilon).
\end{equation}

Usually\cite{peskin} $a=-1/2$ such that $x$ is the midpoint of the two split points. However, there is no \emph{principle} that demands $a$ to be $-1/2$, and if this method is to make sense, the final results must be independent of $a$, as is the case for the chiral anomaly \cite{peskin}. (Looking into concrete process of calculating the chiral anomaly\cite{peskin}, it is easy to see that the chiral anomaly only needs expansion to $\mathcal{O}(\epsilon)$ so that the $a$ dependence is of the form $(a+1)-a=1$.) We will see soon that the point-splitting method cannot be applied to calculating transverse anomalies owing to its non-trivial dependence on $a$.

We first use $\left[\gamma^\rho,\frac{1}{2}\sigma^{\mu\nu}\right]=ig^{\mu\rho}\gamma^\nu-ig^{\nu\rho}\gamma^\mu$ to rewrite $\partial^{[\mu}j^{\nu]}$ as :
\begin{equation}\begin{aligned}
&\partial^\mu j^\nu(x)-\partial^\nu j^\mu(x)\\
=&-i\partial_\rho\left(\bar{\psi}(x+(a+1)\epsilon)\left[\gamma^\rho,\frac{1}{2}\sigma^{\mu\nu}\right] e^{ig\int_{x+a\epsilon}^{x+(a+1)\epsilon}dy\cdot A(y)}\psi(x+a\epsilon)\right).
\end{aligned}\end{equation}

Rearranging terms gives:
\begin{equation}\begin{aligned}
&\partial^\mu j^\nu(x)-\partial^\nu j^\mu(x)\\
=&i\left(\bar{\psi}(x+(a+1)\epsilon)\left(\overrightarrow{\partial}_\rho-\overleftarrow{\partial}_\rho\right)\epsilon^{\mu\nu\rho\sigma}\gamma_\sigma\gamma_5\psi(x+a\epsilon)\right)e^{ig\int_{x+a\epsilon}^{x+(a+1)\epsilon}dy\cdot A(y)}\\
&-i\left(\bar{\psi}(x+(a+1)\epsilon)\overleftarrow{\partial}_\rho\gamma^\rho\sigma^{\mu\nu}\psi(x+a\epsilon)\right) e^{ig\int_{x+a\epsilon}^{x+(a+1)\epsilon}dy\cdot A(y)}\\
&+i\left(\bar{\psi}(x+(a+1)\epsilon)\sigma^{\mu\nu}\gamma^\rho\overrightarrow{\partial}_\rho \psi(x+a\epsilon)\right)e^{ig\int_{x+a\epsilon}^{x+(a+1)\epsilon}dy\cdot A(y)}\\
&+g\bar{\psi}(x+(a+1)\epsilon)\left[\gamma^\rho,\frac{1}{2}\sigma^{\mu\nu}\right] e^{ig\int_{x+a\epsilon}^{x+(a+1)\epsilon}dy\cdot A(y)}\psi(x+a\epsilon)\\
&\times\left(\epsilon^\sigma\partial_\rho A_\sigma(x)+\frac{(a+1)^2-a^2}{2}\epsilon^\sigma\epsilon^\lambda\partial_\lambda\partial_\rho A_\sigma(x)+\frac{(a+1)^3-a^3}{6}\epsilon^\sigma\epsilon^\lambda\epsilon^\kappa\partial_\lambda\partial_\kappa\partial_\rho A_\sigma(x)+\mathcal{O}(\epsilon^4)\right)\\
\end{aligned}\end{equation}

We then use equations of motion for the massless (for simplicity) fermion $\slashed{\overrightarrow{D}}\psi(x)=0$ and $\bar{\psi}(x)\slashed{\overleftarrow{D}}=0$ to get
\begin{equation}\begin{aligned}
&\partial^\mu j^\nu(x)-\partial^\nu j^\mu(x)\\
=&i\left(\bar{\psi}(x+(a+1)\epsilon)\left(\overrightarrow{\partial}_\rho-\overleftarrow{\partial}_\rho\right)\epsilon^{\mu\nu\rho\sigma}\gamma_\sigma\gamma_5\psi(x+a\epsilon)\right)e^{ig\int_{x+a\epsilon}^{x+(a+1)\epsilon}dy\cdot A(y)}\\
&-g\bar{\psi}(x+(a+1)\epsilon)A(x+(a+1)\epsilon)\gamma^\rho\sigma^{\mu\nu}\psi(x+a\epsilon) e^{ig\int_{x+a\epsilon}^{x+(a+1)\epsilon}dy\cdot A(y)}\\
&-g\bar{\psi}(x+(a+1)\epsilon)\sigma^{\mu\nu}\gamma^\rho A(x+a\epsilon)\psi(x+a\epsilon)e^{ig\int_{x+a\epsilon}^{x+(a+1)\epsilon}dy\cdot A(y)}\\
&+g\bar{\psi}(x+(a+1)\epsilon)\left[\gamma^\rho,\frac{1}{2}\sigma^{\mu\nu}\right] e^{ig\int_{x+a\epsilon}^{x+(a+1)\epsilon}dy\cdot A(y)}\psi(x+a\epsilon)\\
&\times\left(\epsilon^\sigma\partial_\rho A_\sigma(x)+\frac{(a+1)^2-a^2}{2}\epsilon^\sigma\epsilon^\lambda\partial_\lambda\partial_\rho A_\sigma(x)+\frac{(a+1)^3-a^3}{6}\epsilon^\sigma\epsilon^\lambda\epsilon^\kappa\partial_\lambda\partial_\kappa\partial_\rho A_\sigma(x)+\mathcal{O}(\epsilon^4)\right)\\
\end{aligned}\end{equation}

We next expand $A_\mu$ at $x$ to $\mathcal{O}(\epsilon^4)$ :
\begin{equation}\begin{aligned}
&\partial^\mu j^\nu(x)-\partial^\nu j^\mu(x)\\
=&i\left(\bar{\psi}(x+(a+1)\epsilon)\left(\overrightarrow{D}_\rho-\overleftarrow{D}_\rho\right)\epsilon^{\mu\nu\rho\sigma}\gamma_\sigma\gamma_5\psi(x+a\epsilon)\right)e^{ig\int_{x+a\epsilon}^{x+(a+1)\epsilon}dy\cdot A(y)}\\
&-g\bar{\psi}(x+(a+1)\epsilon)\gamma^\rho\sigma^{\mu\nu}\psi(x+a\epsilon) e^{ig\int_{x+a\epsilon}^{x+(a+1)\epsilon}dy\cdot A(y)}\\
&\times\left((a+1)\epsilon^\sigma\partial_\sigma A_\rho(x)+\frac{(a+1)^2}{2}\epsilon^\sigma\epsilon^\lambda\partial_\lambda\partial_\sigma A_\rho(x)+\frac{(a+1)^3}{6}\epsilon^\sigma\epsilon^\lambda\epsilon^\kappa\partial_\lambda\partial_\kappa\partial_\sigma A_\rho(x)+\mathcal{O}(\epsilon^4)\right)\\
&-g\bar{\psi}(x+(a+1)\epsilon)\sigma^{\mu\nu}\gamma^\rho\psi(x+a\epsilon) e^{ig\int_{x+a\epsilon}^{x+(a+1)\epsilon}dy\cdot A(y)}\\
&\times\left(a\epsilon^\sigma\partial_\sigma A_\rho(x)+\frac{a^2}{2}\epsilon^\sigma\epsilon^\lambda\partial_\lambda\partial_\sigma A_\rho(x)+\frac{a^3}{6}\epsilon^\sigma\epsilon^\lambda\epsilon^\kappa\partial_\lambda\partial_\kappa\partial_\sigma A_\rho(x)+\mathcal{O}(\epsilon^4)\right)\\
&+g\bar{\psi}(x+(a+1)\epsilon)\left[\gamma^\rho,\frac{1}{2}\sigma^{\mu\nu}\right] e^{ig\int_{x+a\epsilon}^{x+(a+1)\epsilon}dy\cdot A(y)}\psi(x+a\epsilon)\\
&\times\left(\epsilon^\sigma\partial_\rho A_\sigma(x)+\frac{(a+1)^2-a^2}{2}\epsilon^\sigma\epsilon^\lambda\partial_\lambda\partial_\rho A_\sigma(x)+\frac{(a+1)^3-a^3}{6}\epsilon^\sigma\epsilon^\lambda\epsilon^\kappa\partial_\lambda\partial_\kappa\partial_\rho A_\sigma(x)+\mathcal{O}(\epsilon^4)\right).
\end{aligned}\end{equation}

We finally take the $\epsilon\rightarrow0$ limit. From\cite{peskin,he2001}, we have \footnote{Here we only need $\mathcal{O}(A^0)$ of $\left<\psi(x+a\epsilon)\bar{\psi}(x+(a+1)\epsilon)\right>$ because $C$ parity of $j^\mu$ and $A^\mu$ are both odd, and $\mathcal{O}(A^2)$ of $\left<\psi(x+a\epsilon)\bar{\psi}(x+(a+1)\epsilon)\right>$ is of $\mathcal{O}(\epsilon)$ and thus does not make a contribute.} :
\begin{equation}\begin{aligned}
&\left<\psi(x+a\epsilon)\bar{\psi}(x+(a+1)\epsilon)\right>=\frac{i}{2\pi^2}\frac{\gamma^\alpha\epsilon_\alpha}{\epsilon^4}+\mathcal{O}(A^1),\ \lim_{\epsilon\rightarrow0}\frac{\epsilon^\mu\epsilon^\nu}{\epsilon^2}=\frac{1}{4}g^{\mu\nu},\\
&\lim_{\epsilon\rightarrow0}\frac{\epsilon^\mu\epsilon^\nu\epsilon^\rho\epsilon^\sigma}{\epsilon^4}=\frac{1}{24}\left(g^{\mu\nu}g^{\rho\sigma}+g^{\mu\rho}g^{\nu\sigma}+g^{\mu\sigma}g^{\nu\rho}\right).
\end{aligned}\end{equation}

Therefore, the final result (where the Bianchi identity is used and care is taken with Fermi-statistics) is
\begin{equation}\begin{aligned}
&\partial^\mu j^\nu(x)-\partial^\nu j^\mu(x)\\
=&i\bar{\psi}(x)\left(\overrightarrow{D}_\rho-\overleftarrow{D}_\rho\right)\epsilon^{\mu\nu\rho\sigma}\gamma_\sigma\gamma_5\psi(x)\\
&+\frac{g}{\pi^2}\frac{1}{\epsilon^2}F^{\mu\nu}(x)+\frac{g\left[(a+1)^3-a^3\right]}{72\pi^2}\left(2\partial^\rho\partial_\rho F^{\mu\nu}(x)+2\partial^\nu\partial_\rho F^{\mu\rho}(x)-2\partial^\mu\partial_\rho F^{\nu\rho}(x)\right)\\
=&2\bar{\psi}(x)\epsilon^{\mu\nu\rho\sigma}\gamma_\sigma\gamma_5iD_\rho\psi(x)\\
&+\frac{g}{\pi^2}\frac{1}{\epsilon^2}F^{\mu\nu}(x)+\frac{g\left[(a+1)^3-a^3\right]}{18\pi^2}\partial^{\rho}\partial_\rho F^{\mu\nu}(x).
\end{aligned}\end{equation}

This result is not only affected by quadratic divergence\footnote{Unlike the case in Appendix \ref{app:pv}, it seems here that we cannot find a proper way to subtract this divergence because the quadratic divergence of $j^\mu$ is not shown explicitly.}, but also dependent on $a$ nontrivially. The point-splitting method is thus not suitable for transverse anomalies.

\section{Details of Dimensional Renormalization in Section \ref{sec:derivation-DR}}\label{app:dim}

This appendix presents a note for one-loop calculation in dimensional renormalization and an analysis for gauge invariance of the coefficient in (\ref{eq:coeffi-DR}) including that of transverse anomalies.

\begin{figure}[ht]
\centering
\includegraphics[width=6cm]{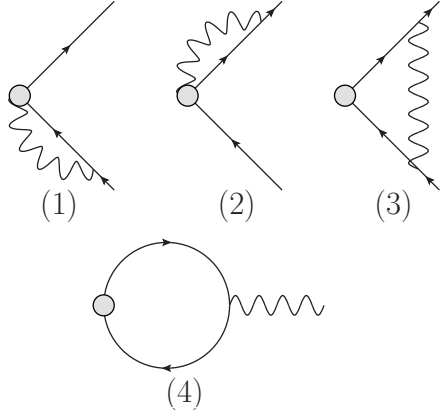}
\caption{One-loop diagrams of $\left<TN\left[\bar{\psi}\bar{\sigma}^{\mu\nu}\gamma^\rho iD_\rho\psi\right]\tilde{\psi}\left(\bar{p}_1\right)\tilde{\bar{\psi}}\left(\bar{p}_2\right)\right>^{\mathrm{prop}}$ and $\left<TN\left[\bar{\psi}\bar{\sigma}^{\mu\nu}\gamma^\alpha iD_\alpha\psi\right]\tilde{A}^\rho(\bar{q})\right>^{\mathrm{prop}}$}
\label{fig:dim1}
\end{figure}

To determine coefficients in tWTI (\ref{tWTI-DR2}) on one-loop order, it is not necessary to use all the algebra in (\ref{eq:coeffi-DR}), and it is more convenient and simple to calculate $\left<TN\left[\bar{\psi}\bar{\sigma}^{\mu\nu}\hat{\gamma}^\rho iD_\rho\psi\right]\tilde{\psi}\left(\bar{p}_1\right)\tilde{\bar{\psi}}\left(\bar{p}_2\right)\right>^{\mathrm{prop}}$ and $\left<TN\left[\bar{\psi}\bar{\sigma}^{\mu\nu}\hat{\gamma}^\alpha iD_\alpha\psi\right]\tilde{A}^\rho(\bar{q})\right>^{\mathrm{prop}}$ (instead of the Green's functions with $N\left[\bar{\psi}\bar{\sigma}^{\mu\nu}\hat{\gamma}^\rho iD_\rho\psi\right]$ replaced by $N\left[\check{g}_{\rho\sigma}\left(\bar{\psi}\bar{\sigma}^{\mu\nu}\gamma^\rho iD^\sigma\psi\right)\right]$) to read out $(b,c,f,r,s)=(b',c',f',r',s')/(1-a)$ directly from (\ref{eq:evan-1}). (Of course, information on $a$ is lost, but this does not matter because $a$ is only an intermediate variable.) Only the following four Feynman diagrams in Fig.\ref{fig:dim1} are relevant.

For the first diagram,
\begin{equation}\begin{aligned}
&\mathscr{E}^{\mu\nu}_{(1)}(\bar{p}_1,\bar{p}_2)\\
=&-2ig^2\int\frac{d^dp}{(2\pi)^d}\bar{\sigma}^{\mu\nu}\left(\hat{\gamma}^\alpha\frac{1}{\slashed{p}+\slashed{\bar{p}}_2-m}\hat{\gamma}_\alpha-(1-\xi)\hat{\slashed{p}}\frac{1}{\slashed{p}+\slashed{\bar{p}}_2-m}\slashed{p}\frac{1}{p^2}\right)\frac{1}{p^2}\\
=&-2ig^2\bar{\sigma}^{\mu\nu}\int\frac{d^dp}{(2\pi)^d}\int_0^1dx\left(\frac{(4-d)\left((1-x)\slashed{\bar{p}}_2-m_0\right)}{\left[-p^2-x(1-x)\bar{p}^2_2+xm_0^2\right]^2}\right.\\
&\left.+(1-\xi)\frac{2(1-x)\hat{p}^2\left(m-\slashed{\bar{p}}_2\right)}{\left[-p^2-x(1-x)\bar{p}^2_2+xm_0^2\right]^3}\right)\\
=&\frac{g^2}{16\pi^2}\bar{\sigma}^{\mu\nu}\left(\slashed{\bar{p}}_2-4m+\left(1-\xi\right)\left(m-\slashed{\bar{p}}_2\right)\right).
\end{aligned}\end{equation}
Similar to the first diagram, we have for the second diagram that
\begin{equation}\begin{aligned}
&\mathscr{E}^{\mu\nu}_{(2)}(\bar{p}_1,\bar{p}_2)\\
=&-2ig^2\int \frac{d^dp}{(2\pi)^d}\left(\hat{\gamma}^\alpha\frac{1}{\slashed{p}+\slashed{\bar{p}}_1-m_0}\hat{\gamma}_\alpha-(1-\xi)\hat{\slashed{p}}\frac{1}{\slashed{p}+\slashed{\bar{p}}_1-m}\slashed{p}\frac{1}{p^2}\right)\bar{\sigma}^{\mu\nu}\frac{1}{p^2}\\
=&-2ig^2\int \frac{d^dp}{(2\pi)^d}\int_0^1dx\left(\frac{(4-d)\left((1-x)\slashed{\bar{p}}_1-m\right)}{\left[-p^2-x(1-x)\bar{p}^2_1+xm^2\right]^2}\right.\\
&\left.+(1-\xi)\frac{2(1-x)\hat{p}^2\left(m-\slashed{\bar{p}}_1\right)}{\left[-p^2-x(1-x)\bar{p}^2_1+xm^2\right]^3}\right)\bar{\sigma}^{\mu\nu}\\
=&\frac{g^2}{16\pi^2}\left(2\slashed{\bar{p}}_1-4m+\left(1-\xi\right)\left(m-\slashed{\bar{p}}_1\right)\right)\bar{\sigma}^{\mu\nu}.
\end{aligned}\end{equation}
The third diagram is a little complicated but still straightforward to calculate :
\begin{equation}\begin{aligned}
&\mathscr{E}^{\mu\nu}_{(3)}(\bar{p}_1,\bar{p}_2)\\
=&2ig^2\int\frac{d^dp}{(2\pi)^d}\left(\gamma^\alpha\frac{1}{\slashed{p}+\slashed{\bar{p}}_1-m}\hat{\slashed{p}}\bar{\sigma}^{\mu\nu}\frac{1}{\slashed{p}+\slashed{\bar{p}}-m}\gamma_\alpha\right.\\
&\left.-(1-\xi)\slashed{p}\frac{1}{\slashed{p}+\slashed{\bar{p}}_1-m}\hat{\slashed{p}}\bar{\sigma}^{\mu\nu}\frac{1}{\slashed{p}+\slashed{\bar{p}}-m}\slashed{p}\frac{1}{p^2}\right)\frac{1}{p^2}\\
=&-2ig^2\int\frac{d^dp}{(2\pi)^d}\cdot 2\int_0^1dx\int_0^xdy\\
&\times\left(\frac{\gamma^\alpha\left(\slashed{p}+(1-y)\slashed{\bar{p}}_1-(1-x)\slashed{\bar{p}}_2+m\right)\hat{\slashed{p}}\bar{\sigma}^{\mu\nu}\left(\slashed{p}-y\slashed{\bar{p}}_1+x\slashed{\bar{p}}_2+m\right)\gamma_\alpha}{\left[-p^2-y(1-y)\bar{p}^2_1-x(1-x)\bar{p}^2_2+2y(1-x)\bar{p}_1\cdot\bar{p}_2+(1-x+y)m^2\right]^3}\right.\\
&+3(1-\xi)(x-y)\left(\slashed{p}-y\slashed{\bar{p}}_1-(1-x)\slashed{\bar{p}}_2\right)\\
&\left.\times\frac{\left(\slashed{p}+(1-y)\slashed{\bar{p}}_1-(1-x)\slashed{\bar{p}}_2+m\right)\hat{\slashed{p}}\bar{\sigma}^{\mu\nu}\left(\slashed{p}-y\slashed{\bar{p}}_1+x\slashed{\bar{p}}_2+m\right)\left(\slashed{p}-y\slashed{\bar{p}}_1-(1-x)\slashed{\bar{p}}_2\right)}{\left[-p^2-y(1-y)\bar{p}^2_1-x(1-x)\bar{p}^2_2+2y(1-x)\bar{p}_1\cdot\bar{p}_2+(1-x+y)m^2\right]^4}\right)\\
=&\frac{g^2}{16\pi^2}\left(-\frac{2}{3}\slashed{\bar{p}}_1\bar{\sigma}^{\mu\nu}+\frac{4}{3}\slashed{\bar{p}}_2\bar{\sigma}^{\mu\nu}+\frac{4}{3}\bar{\sigma}^{\mu\nu}\slashed{\bar{p}}_1-\frac{2}{3}\bar{\sigma}^{\mu\nu}\slashed{\bar{p}}_2\right.\\
&\left.(1-\xi)\left(-2m\bar{\sigma}^{\mu\nu}+\slashed{\bar{p}}_1\bar{\sigma}^{\mu\nu}+\bar{\sigma}^{\mu\nu}\slashed{\bar{p}}_2\right)\right).
\end{aligned}\end{equation}
We now come to the last but most simple diagram :
\begin{equation}\begin{aligned}
&\mathscr{E}^{\mu\nu\rho}_{(4)}(\bar{q})\\
=&-2ig\int\frac{d^dp}{(2\pi)^d}\mathrm{tr}\left[\bar{\gamma}^\rho\frac{1}{\slashed{p}+\slashed{\bar{q}}-m}\hat{\slashed{p}}\bar{\sigma}^{\mu\nu}\frac{1}{\slashed{p}-m}\right]\\
=&8g\left(\bar{q}^\mu\bar{g}^{\nu\rho}-\bar{q}^\nu\bar{g}^{\mu\rho}\right)\int\frac{d^dp}{(2\pi)^d}\int_0^1dx\frac{\hat{p}^2}{\left[-p^2-x(1-x)\bar{q}^2+m^2\right]^2}\\
=&-\frac{g^2}{2\pi^2}\left(i\bar{q}^\mu\bar{g}^{\nu\rho}-i\bar{q}^\nu\bar{g}^{\mu\rho}\right)\left(-\frac{1}{6}\bar{q}^2+m^2\right).
\end{aligned}\end{equation}

When adding these terms together, all the gauge dependent terms cancel out, which verifies our conclusion drawn in Sec.\ref{sec:derivation-DR} that all coefficients in the tWTI (\ref{tWTI-DR2}) are gauge independent to one-loop order. We get these coefficients on one-loop order :
\begin{equation}
b=0,\ c=\frac{g^2}{6\pi^2},\ f=\frac{g^2}{4\pi^2},\ r=\frac{g}{24\pi^2},\ s=\frac{g}{4\pi^2}.
\end{equation}

As for gauge invariance of these coefficients to all orders, some general conclusions drawn in a similar treatment of the chiral anomaly by Bonneau\cite{bonneau2} are enough. We here only quote the contents for the reader's convenience (but with our notations).

The starting point is provided by the action principle ($N\left[\mathcal{O}(x)\right]$ is assumed to be any formally gauge invariant operator) :
\begin{equation}
\frac{\partial}{\partial\xi}\left<TN\left[\mathcal{O}(x)\right]X\right>=\left<T\int d^4yN\left[\frac{i}{2\xi^2}\left(\partial_\mu A^\mu (y)\right)\right]N\left[\mathcal{O}(x)\right]X\right>,
\label{eq:gauge-var1}\end{equation}
where $X\equiv\prod_{i=1}^N\psi\left(x_i\right)\prod_{j=1}^N\bar{\psi}\left(y_j\right)\prod_{k=1}^LA^{\mu_k}\left(z_k\right)$.

Through repeated use of the gauge WTI\footnote{$X\backslash A_{\mu_k}\left(z_k\right)$ means $X$ with $A_{\mu_k}\left(z_k\right)$ stripped away.}, we have
\begin{equation}\begin{aligned}
\left<T\partial_\mu A^\mu(x)N\left[\mathcal{O}(y)\right]X\right>=&-\xi\sum_{k=1}^L\partial_{\mu_k}^xD\left(x-z_k\right)\left<TN\left[\mathcal{O}(y)\right]X\backslash A_{\mu_k}\left(z_k\right)\right>\\
&+ig\xi\sum_{i=1}^N\left(D\left(x-x_i\right)-D\left(x-y_i\right)\right)\left<TN\left[\mathcal{O}(y)\right]X\right>,
\end{aligned}\end{equation}
and we can recast (\ref{eq:gauge-var1}) to many useful forms. We may only focus on the gauge variance of the proper part of Green's functions with $X=A^\rho$ and $X=\psi\left(y\right)\bar{\psi}\left(z\right)$, because expression (\ref{eq:coeffi-DR}) for coefficients in the tWTI only considers these two cases.

Gauge variance of Green's functions is not the focus of this paper, and we thus only quote two main results of \cite{bonneau2} to illustrate the gauge invariance of coefficients in the tWTI. The first result is (B.10) in \cite{bonneau2}, for formally gauge invariant $N\left[\mathcal{O}(x)\right]$
\begin{equation}
\frac{\partial}{\partial\xi}\left<TN\left[\mathcal{O}(x)\right]\prod_{k=1}^LA^{\mu_k}\left(z_k\right)\right>^{\mathrm{prop}}=0.
\end{equation}
This is also established for the non-overall subtracted Green function (see the first sentence after (B.11) in \cite{bonneau2}). Thus $\frac{\partial}{\partial \xi}r'=0$ and $\frac{\partial}{\partial \xi}s'=0$ are simply special cases where $\mathcal{O}=\check{g}_{\rho\sigma}\left(\bar{\psi}\bar{\sigma}^{\mu\nu}\gamma^\rho iD^\sigma\psi\right)$ and $L=1$.

The second result deals with the gauge invariance of $\mathrm{r.s.p.}\overline{\left<TN\left[\mathcal{O}(x)\right]\tilde{\psi}\left(p\right)\tilde{\bar{\psi}}\left(q\right)\right>^{\mathrm{prop}}}$. Figure 3 (which provides a diagrammatical representation of the gauge variance of $\left<TN\left[\mathcal{O}(x)\right]\tilde{\psi}\left(p\right)\tilde{\bar{\psi}}\left(q\right)\right>^{\mathrm{prop}}$) and  Lemma 3 in \cite{bonneau2} (i.e., (B.13.a) and (B.13.b) therein) indicates that, if $\left<TN\left[\mathcal{O}(x)\right]\tilde{\psi}\left(p\right)\tilde{\bar{\psi}}\left(q\right)\right>$ has no trivial part (where a non-trivial diagram was defined by \cite{bonneau2} to be a graph with at least one loop), then 
\begin{equation}
\frac{\partial}{\partial\xi}\mathrm{r.s.p.}\overline{\left<TN\left[\mathcal{O}(x)\right]\tilde{\psi}\left(p\right)\tilde{\bar{\psi}}\left(q\right)\right>^{\mathrm{prop}}}=0.
\end{equation}
Obviously $\left<TN\left[\check{g}_{\rho\sigma}\left(\bar{\psi}\bar{\sigma}^{\mu\nu}\gamma^\rho iD^\sigma\psi\right)(x)\right]\tilde{\psi}\left(p\right)\tilde{\bar{\psi}}\left(q\right)\right>$ has no trivial part owing to the presence of $\check{g}_{\rho\sigma}=\hat{g}_{\rho\sigma}/(d-4)$ (which we take to be zero after finishing all loop integrals), and obtaining the gauge independence of $a,b',c'f'$ is thus straightforawrd.

We get the gauge independence of $b,c,f,r,s$ by combining these two results.

\end{document}